\documentclass[paper]{ieice}
\usepackage[dvipdfmx]{graphicx,xcolor}
\usepackage[fleqn]{amsmath}
\usepackage{newtxtext}
\usepackage[varg]{newtxmath}
\usepackage{subfigure}
\usepackage{cite}
\usepackage{comment}
\usepackage{hhline}

\field{A}
\vol{103}
\no{12}
\SpecialSection{Smart Multimedia \& Communication Systems}
\theme{}
\title{A Reversible Data Hiding Method in Compressible Encrypted Images}
\authorlist{%
 \authorentry[imaizumi@chiba-u.jp]{Shoko IMAIZUMI}{m}{cu1}\MembershipNumber{0215457}
 \authorentry{Yusuke IZAWA}{n}{cu2}
 \authorentry{Ryoichi HIRASAWA}{n}{cu2}
 \authorentry[kiya@sd.tmu.ac.jp]{Hitoshi KIYA}{f}{tmu}\MembershipNumber{}
}
\affiliate[cu1]{The authors is with the Graduate School of Engineering,
Chiba University, Chiba-shi, 263-8522 Japan.
}
\affiliate[cu2]{The authors is with the Graduate School of Science and Engineering,
Chiba University, Chiba-shi, 263-8522 Japan.
}
\affiliate[tmu]{The author is with the Faculty of System Design, Tokyo
Metropolitan University, Hino-shi, 191-0065 Japan.
}
\received{2020}{2}{7}
\revised{2020}{5}{25}

\begin{document}
\maketitle
\begin{summary}
 We propose a reversible data hiding (RDH) method in compressible encrypted images called the encryption-then-compression (EtC) images. %, called encryption-then-compression (EtC) images. 
 The proposed method allows us to not only embed a payload in encrypted images but also compress the encrypted images containing the payload. 
 In addition, the proposed RDH method can be applied to both plain images and encrypted ones, and the payload can be extracted flexibly in the encrypted domain or from the decrypted images. 
 Various RDH methods have been studied in the encrypted domain, but they are not considered to be two-domain data hiding, and the resultant images cannot be compressed by using image coding standards, such as JPEG-LS and JPEG 2000.
 In our experiment, the proposed method shows high performance in terms of lossless compression efficiency by using JPEG-LS and JPEG 2000, data hiding capacity, and marked image quality. 
\end{summary}

\section{Introduction}
\label{sec:1}
 Reversible data hiding (RDH) can perfectly retrieve original images from the marked images. 
 It is particularly useful for medical, military, evidential images, and so forth~\cite{IEEE-A2016:YQShi, IEEE-T2006:ZNi, IEEE-T2007:DMThodi, ICIP2007:MFujiyoshi}.  
 For secure sharing of secret images, the image owner may encrypt the images beforehand. 
 Here, we assume that some images are encrypted to protect their visual information. 
 A third party, such as the system/channel administrator, may append additional information to the encrypted images. 
 The image user hopes to restore the high-quality image after decryption even though some additional information is still contained in the image. 
 It is further helpful to perfectly restore the original image by extracting the information if needed.
 Hence, RDH methods for encrypted images have been studied~\cite{IEEE-SPL2011:XZhang, IEEE-SPL2012:WHong, Springer-MSSP2018:LXiong, IEEE-T2012:XZhang, IEEE-T2013:KMa, SCN2013:XZhang}. 
 By using those methods, the payload embedded in the encrypted domain can be extracted from the decrypted image. 
 Conversely, in the case that a payload is embedded in the plain domain and can be extracted from the encrypted image, image retrieval in the encrypted domain can be attained.  
 Those integration methods of reversible data hiding and encryption consequently give three kinds of user authority: (i) decryption only, (ii) data extraction only, and (iii) both decryption and data extraction. 
 Further, if two different payloads are embedded in the plain and encrypted domains independently and could be extracted from either domain, the use of the application can be further extended. 
 For instance, we assume that some images are preliminarily encrypted by the owner and then the encrypted images are outsourced to a third party. 
 The owner would also like the encrypted images to be searchable without decryption. 
 In such a scenario, it would be helpful if the owner could embed the image information such as the ownership and image content in the plain domain while the third party could embed additional information, e.g., the server information and time stamps, in the encrypted domain. 
 We propose a flexible RDH method in both a single domain and two domains.  

 Zhang~\cite{IEEE-SPL2011:XZhang} proposed an RDH method in the encrypted domain. 
 This method encrypts the whole image pixel-by-pixel using the exclusive-or operation and thus cannot compress the encrypted image containing a payload (hereafter, output image). 
 The RDH process cannot be performed in two domains. 
 Additionally, the payload may not be correctly extracted, leading to the original image not being retrieved. 
 Hong et al.~\cite{IEEE-SPL2012:WHong} modified Zhang's method~\cite{IEEE-SPL2011:XZhang}, but nevertheless the payload cannot be perfectly extracted and both the compression efficiency and two-domain data hiding are not considered. 
 In Xiong et al.'s method~\cite{Springer-MSSP2018:LXiong}, integer wavelet transform is introduced before encryption to attain a higher embedding rate and higher PSNR with the same amount of payload.
 It has been confirmed that this method can extract the payload and restore the original image perfectly under certain conditions. 
 The reversibility is not ensured under different conditions. 
 Moreover, this method does not consider compression of output images and cannot be applied to the two-domain data hiding. 
 In another work~\cite{IEEE-T2012:XZhang}, the quality of the marked image, which is equal to that of the decryption-only image, has been improved compared to \cite{IEEE-SPL2011:XZhang}. 
 However, if the payload amount is too large, the payload cannot be correctly extracted. 
 Further, this method cannot compress the output images by using image coding standards, such as JPEG-LS~\cite{JPEGLS} and JPEG 2000~\cite{JP2}, and cannot be extended to the two-domain data hiding. 
 Ma et al.'s method~\cite{IEEE-T2013:KMa} gives a particular user authority. 
 While generally the original image cannot be retrieved without data extraction, this method can restore the original image after only the decryption.  
 Zhang~\cite{SCN2013:XZhang} proposed another approach that embeds a payload in the plain domain and can extract the payload from the encrypted image. 
 In this method, we may also embed a payload in the encrypted domain and extract the payload from the decrypted image. 
 In those methods~\cite{IEEE-T2013:KMa, SCN2013:XZhang}, however, both the compression of the output images and the two-domain data hiding cannot be achieved.

 In this paper, we propose a flexible RDH method in both a single domain and two domains. 
 The proposed method embeds a payload in the encrypted domain and can extract the payload from the decrypted image.
 Conversely, this method can embed a payload in the plain domain and extract the payload from the encrypted image. 
 Furthermore, the RDH method can be extended to both the plain and encrypted domains in our method. 
 This extension contributes to expanding the types of user authorities.  
 We adopt the compressible encryption (CE) method~\cite{IEICE-T2015:KKurihara, IEICE-T2018:SImaizumi, IEICE-T2017:KKurihara, ICASSP2015:OWatanabe} for high compression efficiency of the output image.
 Additionally, the RDH method based on histogram shift (HS)~\cite{IEEE-T2006:ZNi} is introduced in our method to embed a payload in both the plain and encrypted domains and flexibly extract the payload in either domain. 
 In the proposed method, complex conditions need to be configured to define the data hiding order and the target blocks for encryption. 
 Those conditions enable complete data extraction without any loss of the payload. 
 Through our experiments, we confirm the effectiveness of the proposed method in terms of lossless compression performance using JPEG-LS and JPEG 2000, data hiding capacity/marked-image quality, and robustness against ciphertext-only attacks (COAs).

\section{Preparation}
\label{sec:2}

\subsection{Compressible encryption method}
\label{ssec:2-1}
\begin{figure}[tb]
 \centering

  \includegraphics[width=0.98\columnwidth]{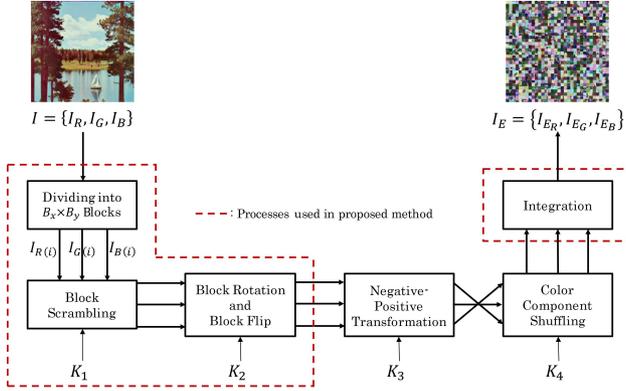}
 \caption{Compressible encryption method for EtC system.}
 \label{fig:ce}
\end{figure}
 Innovative frameworks for secure image transmission were proposed for encryption-then-compression (EtC) systems~\cite{Elsevier_DSP2017:MKumar, IEEE-T2014:JZhou, IEEE-T2010:WLiu, IEEE-T2004:MJohnson, IEICE-T2015:KKurihara, IEICE-T2018:SImaizumi, IEICE-T2017:KKurihara, ICASSP2015:OWatanabe}.
 In the systems, encryption is performed by an image owner before compression/transmission.
 Common-key cryptosystems are frequently used for protecting visual information on plain images. 
 However, most of those methods do not support lossless compression of the encrypted images. 
 The CE method has been developed for a solution to this issue~\cite{IEICE-T2015:KKurihara, IEICE-T2018:SImaizumi, IEICE-T2017:KKurihara, ICASSP2015:OWatanabe}. 
 The images encrypted by using this method are called EtC images.
 We introduce the CE method to our proposed method as an encryption algorithm. 

 The block diagram of the CE method is shown in Fig.~\ref{fig:ce}. 
 We describe the detailed procedure as follows.  

\begin{description}
 \item{\bf{Step 1}:} Divide the original image $I =\{I_R, I_G, I_B\}$
	    with $M \times N$ pixels into multiple blocks with $B_x
	    \times B_y$ pixels.
 \item{\bf{Step 2}:} Scramble the position of each block using a random
	    number generated by key $K_1$.
 \item{\bf{Step 3}:} Rotate and flip each block using a random number
	    generated by key $K_2$.
 \item{\bf{Step 4}:} Perform the negative-positive transformation on each
	    block using a random number generated by key $K_3$.
 \item{\bf{Step 5}:} Shuffle the R, G, and B components in each block
	    using a random number generated by key $K_4$.
 \item{\bf{Step 6}:} Integrate all blocks and generate the encrypted
	    image.
\end{description}

 It is noted that the keys $K_1$, $K_2$, and $K_3$ are commonly used among the three color components in the fundamental CE method~\cite{IEICE-T2015:KKurihara}.
 In another study, those keys are independently used among the three color components in the CE method~\cite{IEICE-T2018:SImaizumi}. 
 In the case above, each key is divided into three elemental keys, such as $K_1 = \{K_{1,R}, K_{1,G}, K_{1,B}\}$.
 Each color component is consequently encrypted by a different key. 
 The proposed method can adopt either case.

 The first two permutation processes in the CE method, which are described in Steps 2 and 3, do not transform the image histogram. 
 We focus on this feature and introduce these processes to our encryption algorithm.  
 
\subsection{Reversible data hiding method based on histogram shift~\cite{IEEE-T2006:ZNi}}
\label{ssec:2-2}
 \begin{figure}[t]
 \centering

  \subfigure[Step 1]{%
    \includegraphics[width=.45\columnwidth]{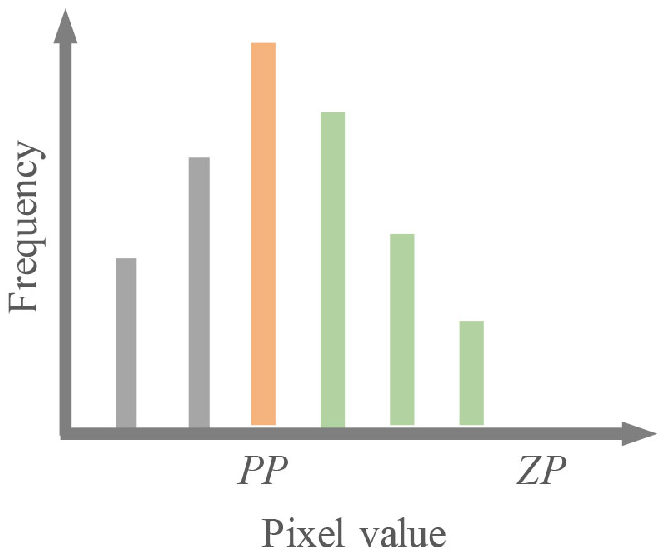}%
    \label{sfig:hs1}%
   }%
   \hfil%
  \subfigure[Step 2]{%
   \includegraphics[width=.45\columnwidth]{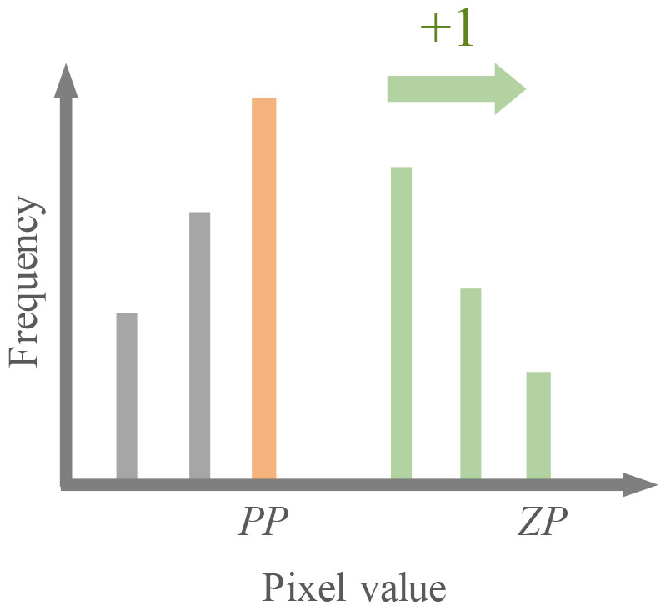}%
   \label{sfig:hs2}%
  }%

  \subfigure[Step 3]{%
    \includegraphics[width=.45\columnwidth]{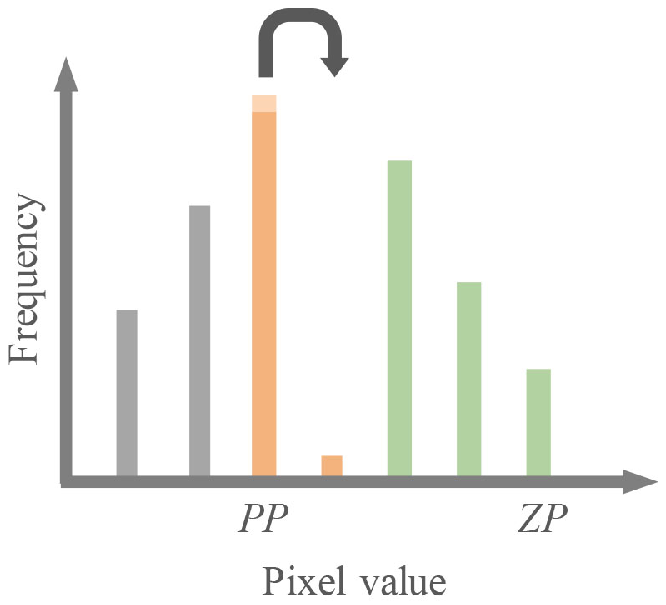}%
    \label{sfig:hs1}%
   }%
   \hfil%
  \subfigure[Step 4]{%
   \includegraphics[width=.45\columnwidth]{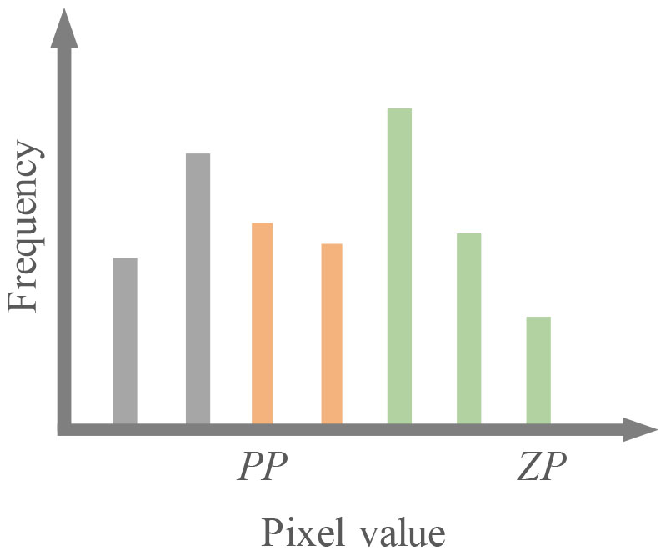}%
   \label{sfig:hs2}%
  }%
 \caption{Procedure of HS-based RDH.}
 \label{fig:hs}
 \end{figure}

 \begin{figure*}[t]
 \centering

  \includegraphics[width=1.95\columnwidth]{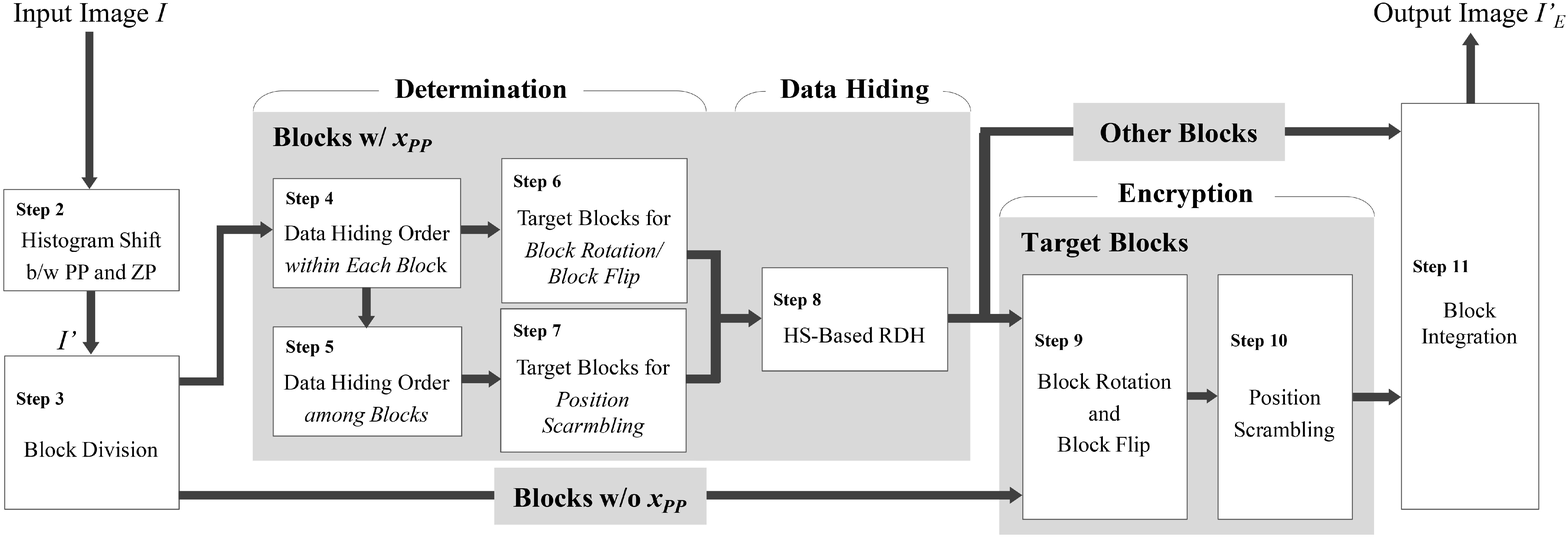}
 \caption{Procedure of proposed method, where payload is embedded in plain domain.}
 \label{fig:flow}
 \end{figure*}

 Ni et al. proposed an RDH method for grayscale images, where the quality of the marked image can be maintained by slightly shifting the image histogram~\cite{IEEE-T2006:ZNi}. 
 Figure~\ref{fig:hs} illustrates the embedding procedure of the HS-based RDH method.
 We describe each step as follows. 
 \begin{description}
 \item[\bf{Step 1}:] A pair of a peak point (hereafter, $PP$), which is the bin with the highest frequency of appearance, and a zero point (hereafter, $ZP$), which is the bin with no pixels, is explored from the image histogram. 
                     If there are multiple pairs of $PP$ and $ZP$, we should take the pair that has the shortest distance between the $PP$ and $ZP$.

 \item[\bf{Step 2}:] We assume that $X$ is the pixel value of pixel $x_X$, where $x_X$ is a pixel in grayscale image $I_g$. If $X$ is located between the $PP$ and $ZP$, it is shifted according to the following equation.
    \begin{eqnarray}
         X' = \begin{cases}
           X + 1, \quad X \in (PP, ZP) ~~~  {\rm if} ~~ PP<ZP \\
           X - 1, \quad X \in (ZP, PP) ~~~  {\rm if} ~~ PP>ZP,
    \end{cases}
    \label{eq:hs1}
  \end{eqnarray}
  where $X'$ denotes the pixel value after shifting. The shifted pixel is depicted as $x_{X'}$. 
  Accordingly, the adjacent bin of $PP$ becomes empty. 
  \item[\bf{Step 3}:] The payload is embedded into pixels $x_{PP}$.
   If the to-be-embedded bit is 1, the pixel value $PP$ is shifted to the empty bin as follows: 
   \begin{eqnarray}
         PP' = \begin{cases}
           PP+1 & {\rm if} ~~ PP<ZP \\
           PP-1 & {\rm if} ~~ PP>ZP,
  \end{cases}
  \label{eq:hs2}
  \end{eqnarray}
  where $PP'$ is the marked pixel value. 
  In contrast, if the to-be-embedded bit is 0, the pixel value is unchanged, that is,
  \begin{equation}
        PP' = PP.
  \label{eq:hs3}
  \end{equation}
  Consequently, $x_{PP}$ turns into marked pixel $x_{PP'}$.

 \item[\bf{Step 4}:] Step 3 is repeated until the whole of the payload is completely embedded.   
 \end{description}
 
 In this method, the data hiding capacity is equal to the total number of pixel $x_{PP}$s. 
 The capacity is increased when the number of $x_{PP}$s is large. 
 If the number of $x_{PP}$s is less than the payload amount, the above steps are repeatedly performed until the whole of the payload is completely embedded. 
 Therefore, a certain bin might be used as the $PP$ more than once. 
 Note that it is necessary to store the $PP$ and $ZP$ values as side information. 
 Those values are generally embedded into the LSBs of the selected pixels. 
 If there is no $ZP$ in the histogram, we adopt the lowest point ($LP$), which is the bin with the lowest frequency of appearance. 
 In that case, the original pixels with the $LP$ should be recorded, and this information is embedded with the pure payload to perfectly retrieve the original image. 
 For data extraction, we perform the above procedure in the opposite order.
 
\section{Proposed method}
\label{sec:3}
 We elaborate our proposed framework to embed a payload in the encrypted domain and extract the payload from the decrypted image. 
 The proposed method can also embed a payload in the plain domain and extract the payload from the encrypted domain.
 Hereafter, we describe the proposed algorithm for the latter case, but it is easy to apply the procedure to the former case by interchanging the steps for encryption and data hiding.
 We first specify the data hiding/encryption processes with the detailed conditions and then delineate the data extraction. 
 Subsequently, an extended framework, where two different payloads can be embedded in both the plain and encrypted domains independently, is described. 

\subsection{Overview of proposed method}
\label{ssec:3-1}
 \begin{description}
  \item{\bf{(A) Embedded in plain domain}}
  
 Figure~\ref{fig:flow} shows the block diagram of the proposed method with data hiding in a single domain. 
 In the proposed method, we use the two permutation processes of the CE method illustrated in Fig.~\ref{fig:ce}, namely, position scrambling and block rotation/flip. 
 The other two processes, that is, negative-positive transformation and color component shuffling, are omitted to ensure the reversibility of our algorithm. 
 Here, we describe the whole procedure including HS-based RDH and CE.
 
 \begin{description}
  \item{\bf{Step 1}:} Explore the $PP$ and $ZP$ in the histogram of original image $I$.
  \item{\bf{Step 2}:} Obtain the intermediate image $I'$ by shifting the histogram between the $PP$ and $ZP$.
                      The shifted pixel values $X'$ are given by Eq.~\eqref{eq:hs1}.
  \item{\bf{Step 3}:} Divide the image $I'$ into multiple blocks with $b_{x} \times b_{y}$ pixels, and define the $a$-th block containing $x_{PP}$ as $B_{x_{PP}}(a)$.
  \item{\bf{Step 4}:} Determine the data hiding order within each block $B_{x_{PP}}(a)$ (details in \ref{sssec:3-1-1}).                      
  \item{\bf{Step 5}:} Determine the data hiding order among the blocks $B_{x_{PP}}(a)$ (details in \ref{sssec:3-1-2}).
  \item{\bf{Step 6}:} Determine the target blocks $B_{r}$ for block rotation and block flip out of all the blocks (details in \ref{sssec:3-1-3}).
  \item{\bf{Step 7}:} Determine the target blocks $B_{p}$ for position scrambling out of all the blocks (details in \ref{sssec:3-1-4}).                                           
  \item{\bf{Step 8}:} Embed a payload into the pixels $x_{PP}$ in the data hiding order defined in Steps 4 and 5. 
                      The marked pixel values $PP'$ are given by Eqs.~\eqref{eq:hs2} and \eqref{eq:hs3}.
  \item{\bf{Step 9}:} Rotate and flip each block $B_{r}$. 
  \item{\bf{Step 10}:} Scramble the position of the blocks $B_{p}$. 
  \item{\bf{Step 11}:} By concatenating all the blocks, obtain the output image $I'_{E}$. 
 \end{description}

  \item{\bf{(B) Embedded in encrypted domain}}
  
 In the case that the payload is embedded in the encrypted domain, the CE process of Steps 9 and 10 is performed previously. 
 Then, the payload is embedded into the EtC image by Step 8. 
 Finally, the output image $I'_{E}$ is obtained in Step 11. 
 Hereafter, we elaborate the conditions for the data hiding order and the target blocks for CE. 
 \end{description}

 \begin{figure}[t]
 \centering

  \subfigure[Exploring $x_{PP}$ from each corner]{%
    \includegraphics[width=.4\columnwidth]{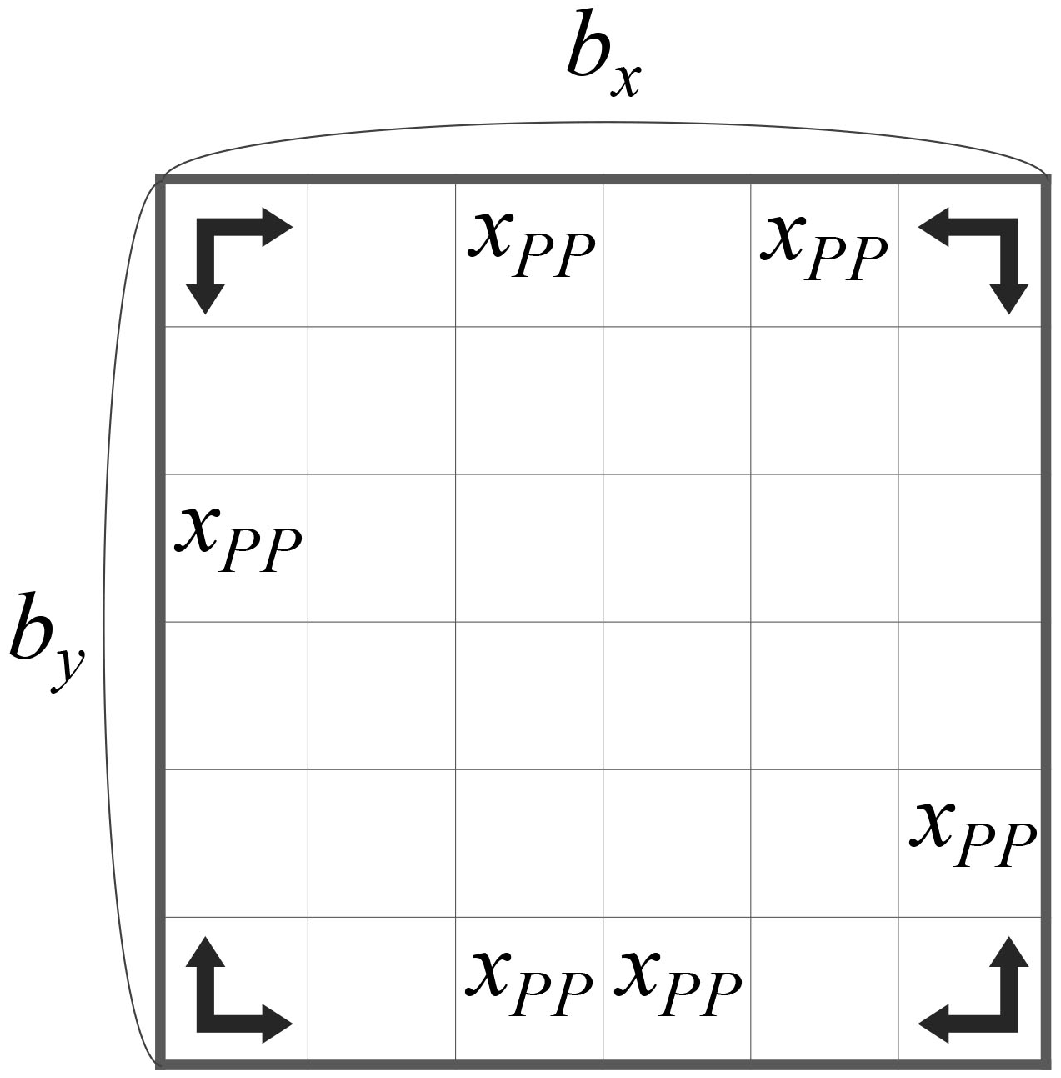}%
    \label{sfig:dho_within1}%
   }%

  \subfigure[Obtaining shortest distances between $i$-th $x_{PP}$ and corner/$i-1$-th $x_{PP}$]{%
    \includegraphics[width=.775\columnwidth]{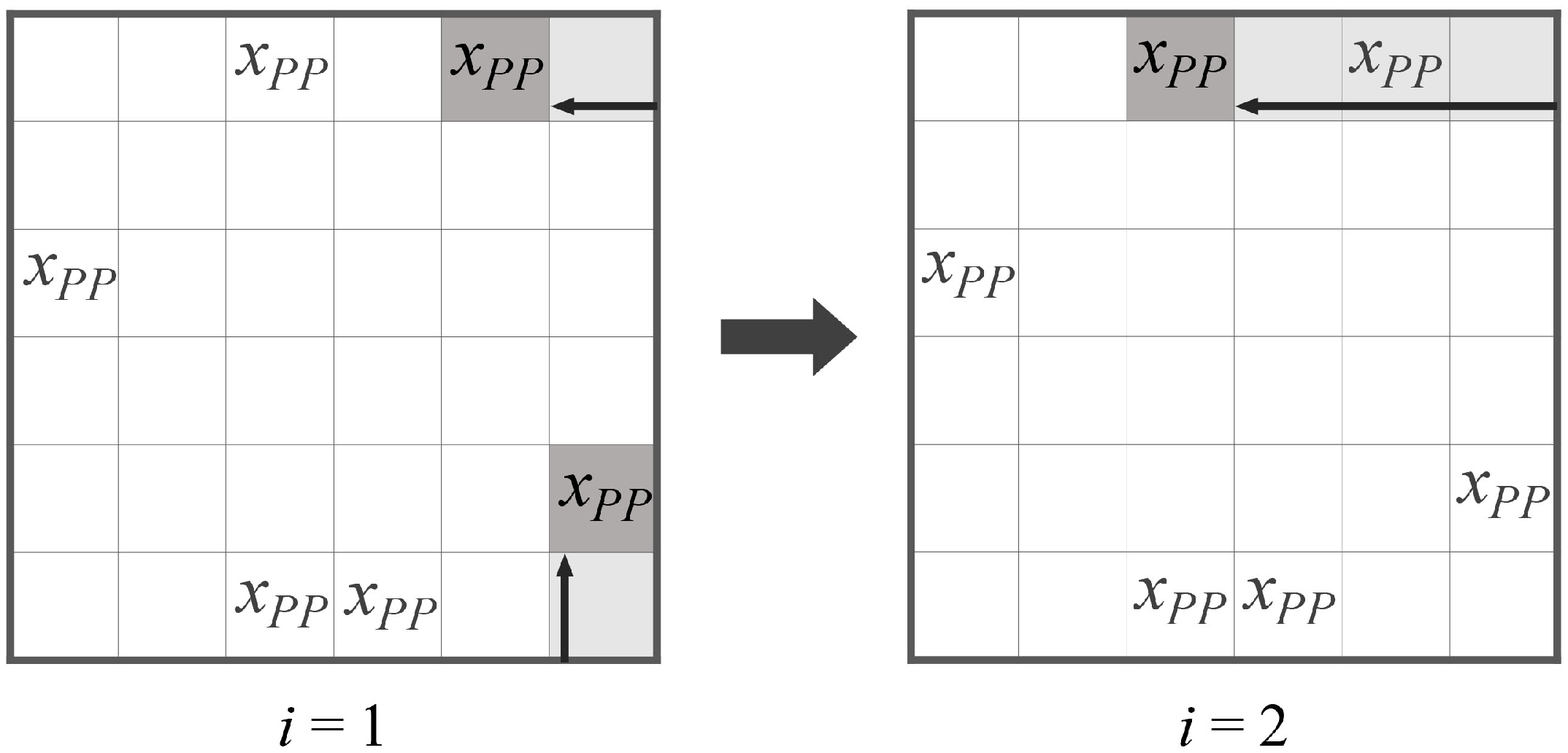}%
    \label{sfig:dho_within2}%
   }%
 \caption{Determination of data hiding order within each block.}
 \label{fig:dho_within}
 \end{figure}
 
\subsubsection{Data hiding order within each block} 
\label{sssec:3-1-1}
 Here, we detail how to determine the data hiding order within each block in Step 4 of \ref{ssec:3-1}.
 The data hiding order in each block is basically equal to the raster scan order. 
 However, we need to consider the block rotation process in encryption for perfect data extraction. 
 We assume that the block $B_{x_{PP}}(a)$ contains $N$ of the pixels $x_{PP}$. 
 The detailed procedure to define the order is described as follows.  

 \begin{description}
  \item{\bf{(1)}} Set $i=1$.
  \item{\bf{(2)}} Explore the $i$-th $x_{PP}$ $(i \in [1, N])$ from each corner in the block $B_{x_{PP}}(a)$, as shown in Fig.~\ref{sfig:dho_within1}, and record the distances between each corner and $x_{PP}$.
  \item{\bf{(3)}} Set the corner with the shortest distance to the $i$-th $x_{PP}$ as the origin/direction of the raster scan for data hiding into $x_{PP}$.
  \item{\bf{(4)}} As shown in Fig.~\ref{sfig:dho_within2}, if there are multiple corners/directions with the shortest distance, $i$ is increased by 1, namely $i=i+1$, and go back to (2).
                 In that case, the distances between the $i-1$-th and $i$-th $x_{PP}$s should be recorded.
  \item{\bf{(5)}} If there are still multiple corners/directions with the shortest distance in the case of $i=N$, the left-top corner and left-to-right direction are chosen as the origin and scanning direction.                 
 \end{description} 

\subsubsection{Data hiding order among blocks} 
\label{sssec:3-1-2}
 The data hiding order among the blocks $B_{x_{PP}}(a)$ in Step 5 of \ref{ssec:3-1} is defined according to the following steps. 

 \begin{description}
  \item{\bf{(1)}} Sort the data hiding blocks in descending order of the number of $x_{PP}$s in each block.
  \item{\bf{(2)}} For the blocks with the same number of $x_{PP}$s, sort the data hiding blocks in ascending order of the number of pixels between the $PP$ and $ZP$ in each block.
  \item{\bf{(3)}} For the blocks with the same number of pixels between the $PP$ and $ZP$, obtain the shortest distance between the origin of the data hiding order within the block and the $j$-th $x_{PP}$ (set $j=1$, $j \in [1, N])$.
                 Sort the data hiding blocks in ascending order of the shortest distance.
                 If there are multiple blocks with the same distance, $j$ is increased by 1, namely $j=j+1$, and repeat this step. 
                 In that case, the distances between the $j-1$-th and $j$-th $x_{PP}$s should be compared. 
  \item{\bf{(4)}} For the blocks with the same distance in (3), sort the data hiding blocks in ascending order of $a$. 
 \end{description}

\subsubsection{Target blocks for block rotation and flip process} 
\label{sssec:3-1-3}
 If the encryption processes are performed on all blocks in the image, the payload cannot be extracted correctly. 
 We need to determine the target blocks for encryption based on several conditions.
 In Step 6 of \ref{ssec:3-1}, the target blocks for the rotation and flip process must satisfy either of the following conditions. 

 \begin{description}
  \item{\bf{(a)}} Blocks $B_{x_{PP}}$, where the data hiding order is determined by (1) -- (4) of \ref{sssec:3-1-1}.
  \item{\bf{(b)}} Blocks that do not contain $x_{PP}$. 
 \end{description} 
 The blocks $B_{x_{PP}}$, where the data hiding order is determined by (5) of \ref{sssec:3-1-1}, are excluded from the target blocks for the rotation and flip process.

\subsubsection{Target blocks for position scrambling process} 
\label{sssec:3-1-4}
  In a similar fashion, the target blocks for the position scrambling process must satisfy either of the following conditions in Step 7 of \ref{ssec:3-1}. 

 \begin{description}
  \item{\bf{(a)}} Blocks $B_{x_{PP}}$, where the data hiding order is determined by (1) -- (3) of \ref{sssec:3-1-2}.
  \item{\bf{(b)}} Blocks that do not contain $x_{PP}$. 
 \end{description} 
 The blocks $B_{x_{PP}}$, where the data hiding order is determined by (4) of \ref{sssec:3-1-2}, are excluded from the target blocks for the position scrambling process.

\subsection{Data extraction}
\label{ssec:3-2}
 \begin{figure*}[t]
 \centering

  \includegraphics[width=1.75\columnwidth]{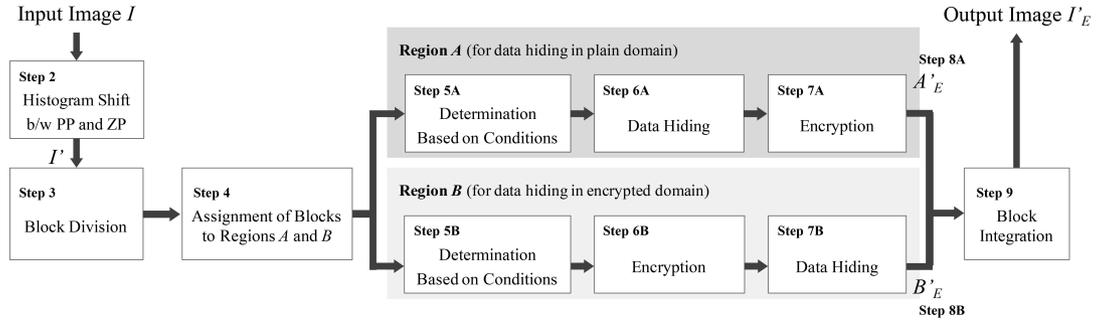}
 \caption{Extended framework for data hiding in two domains.}
 \label{fig:flow_ex}
 \end{figure*}
 
 The payload embedded in \ref{ssec:3-1} can be extracted from the output image $I'_{E}$ without decryption.
 After the block division of $I'_{E}$ and the definition of the data extracting order, the payload bits are extracted from the marked blocks in series.
 The detailed procedure is explained as follows.
 
 \begin{description}
%  \item{\bf{Step 1}:} \textcolor{red}{Explore the marked pixel values $PP'$ in the histogram of the output image $I'_{E}$.
%                      $PP'$ is defined by                      
%                       \begin{eqnarray}
%                        PP' = \begin{cases}
%                        PP~~{\rm or}~~PP+1 & {\rm if} ~~ PP<ZP \\
%                        PP~~{\rm or}~~PP-1 & {\rm if} ~~ PP>ZP.
%                        \end{cases}
%                        \label{eq:hsrev1}
%                       \end{eqnarray}}                  
  \item{\bf{Step 1}:} Divide the image $I'_{E}$ into blocks, and define the $a$-th block containing marked pixel $x_{PP'}$ as $B_{x_{PP'}}(a)$.
 \end{description}
 For all the blocks $B_{x_{PP'}}(a)$,
 \begin{description}
 \item{\bf{Step 2}:} Determine the data extracting order within each block according to the conditions described in \ref{sssec:3-1-1}.
 \item{\bf{Step 3}:} Determine the data extracting order among the blocks according to the conditions described in \ref{sssec:3-1-2}.
 \item{\bf{Step 4}:} According to the data hiding order defined in Steps 2 and 3, extract the payload from $x_{PP'}$. 
                     The original pixels $x_{PP}$ are retrieved as follows.
                     If the embedded bit is 1, the marked pixel value $PP'$ is shifted as 
                     \begin{eqnarray}
                      PP = \begin{cases}
                      PP'-1 & {\rm if} ~~ PP<ZP \\
                      PP'+1 & {\rm if} ~~ PP>ZP.
                      \end{cases}
                      \label{eq:hsrev2}
                     \end{eqnarray}
                     If the embedded bit is 0, the pixel is unchanged, that is,
                     \begin{equation}
                      PP = PP'.
                     \label{eq:hsrev3}
                     \end{equation}
 \end{description}
For all the blocks, 
 \begin{description}
 \item{\bf{Step 5}:} By concatenating all the blocks and shifting the histogram between the $PP$ and $ZP$ according to the following equation, obtain the EtC image $I_{E}$ without the payload.
    \begin{eqnarray}
     \hspace{-20pt}
         X = \begin{cases}
           X' - 1, ~~ X' \in (PP+1, ZP+1) ~~~ {\rm if} ~ PP<ZP \\
           X' + 1, ~~ X' \in (ZP-1, PP-1) ~~~ {\rm if} ~ PP>ZP.
    \end{cases}
    \label{eq:hsrev4}
  \end{eqnarray}
 \end{description}

 In accordance with the above procedure, the payload embedded in the plain domain can be perfectly extracted from the EtC image.
 When the EtC image $I_{E}$ is decrypted, the original image is retrieved completely. 
 In this way, the proposed method can flexibly extract the payload and decrypt the EtC image without regard to the order of extraction and decryption.
 This is due to the invariance of the image histogram before/after encryption.
 Consequently, the receiver types can be classified into three cases: extraction only, decryption only, and both extraction and decryption. 

 In the case that the payload embedded into the EtC image is extracted from the decrypted image, the output image $I'_{E}$ is first decrypted based on the conditions defined in \ref{sssec:3-1-3} and \ref{sssec:3-1-4}.
 Then, the payload is extracted according to the extracting order (Steps 1 -- 4). 
 Finally, the original image is retrieved by Step 5.

\subsection{Data hiding in two domains}
\label{ssec:3-3}
 \begin{figure*}[t]
 \centering

  \subfigure[Image 1]{%
    \includegraphics[width=.475\columnwidth]{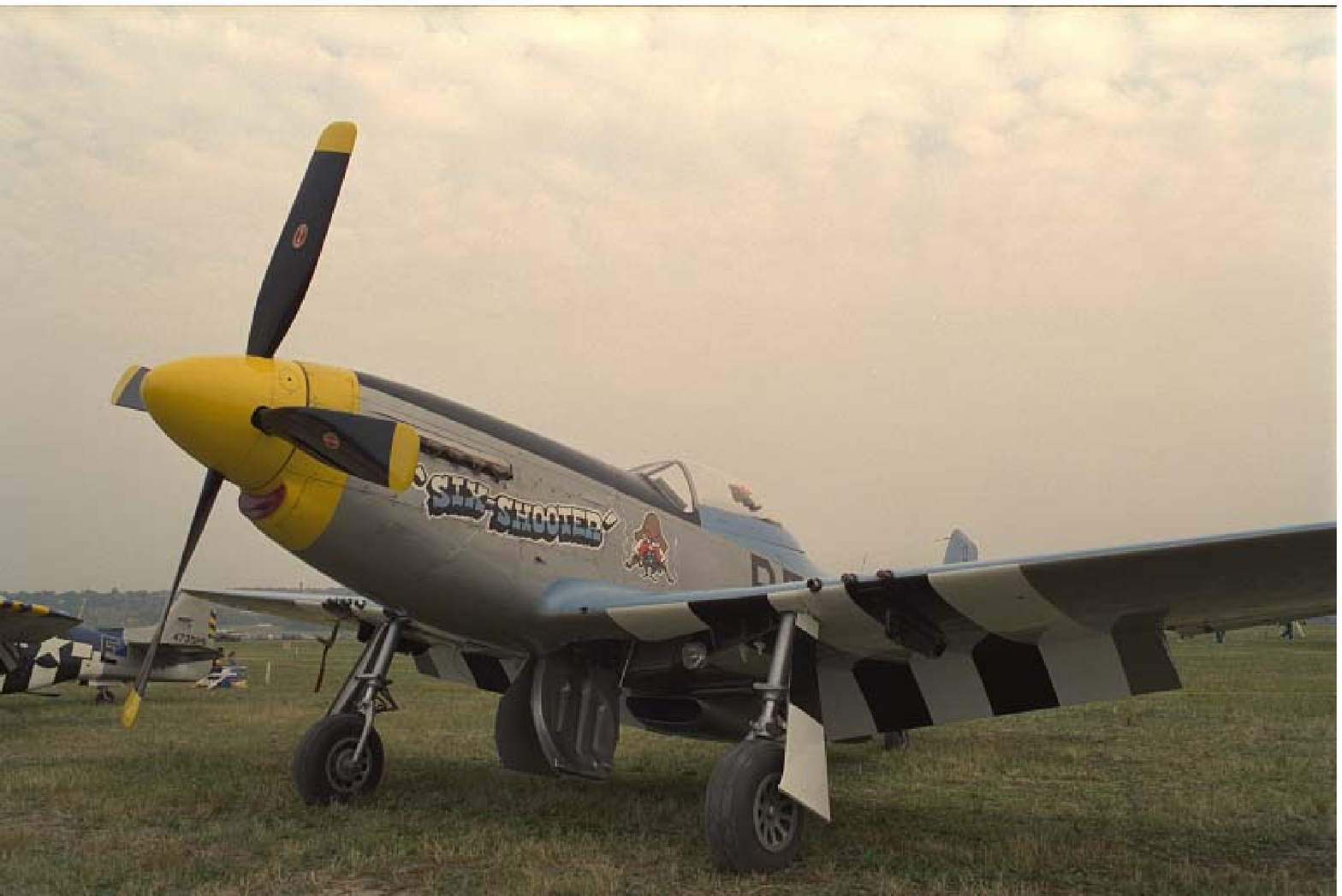}%
    \label{sfig:img1}%
   }%
   \hfil%
  \subfigure[Image 2]{%
   \includegraphics[width=.475\columnwidth]{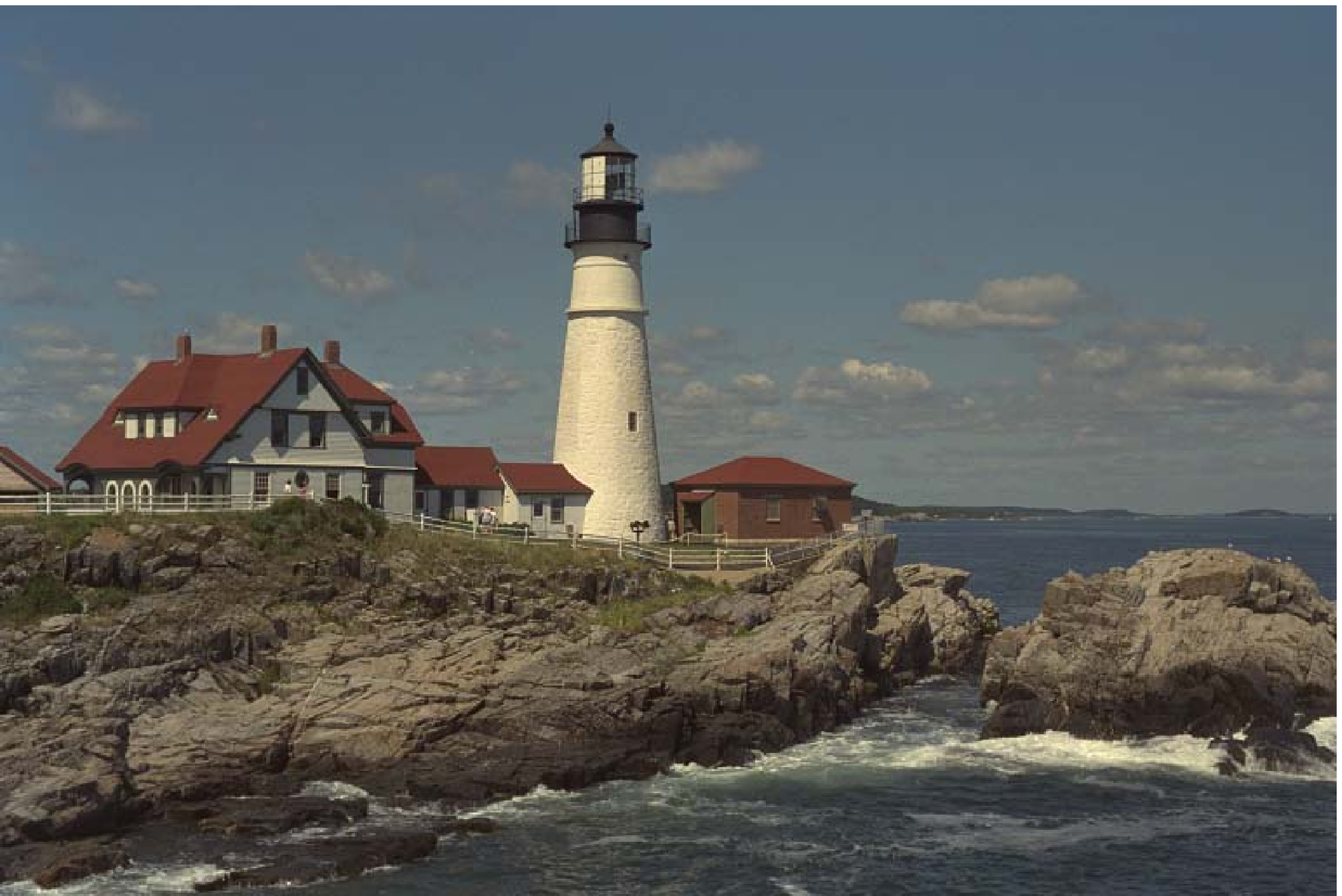}%
   \label{sfig:img2}%
   }%
   \hfil%
  \subfigure[Image 3]{%
    \includegraphics[width=.475\columnwidth]{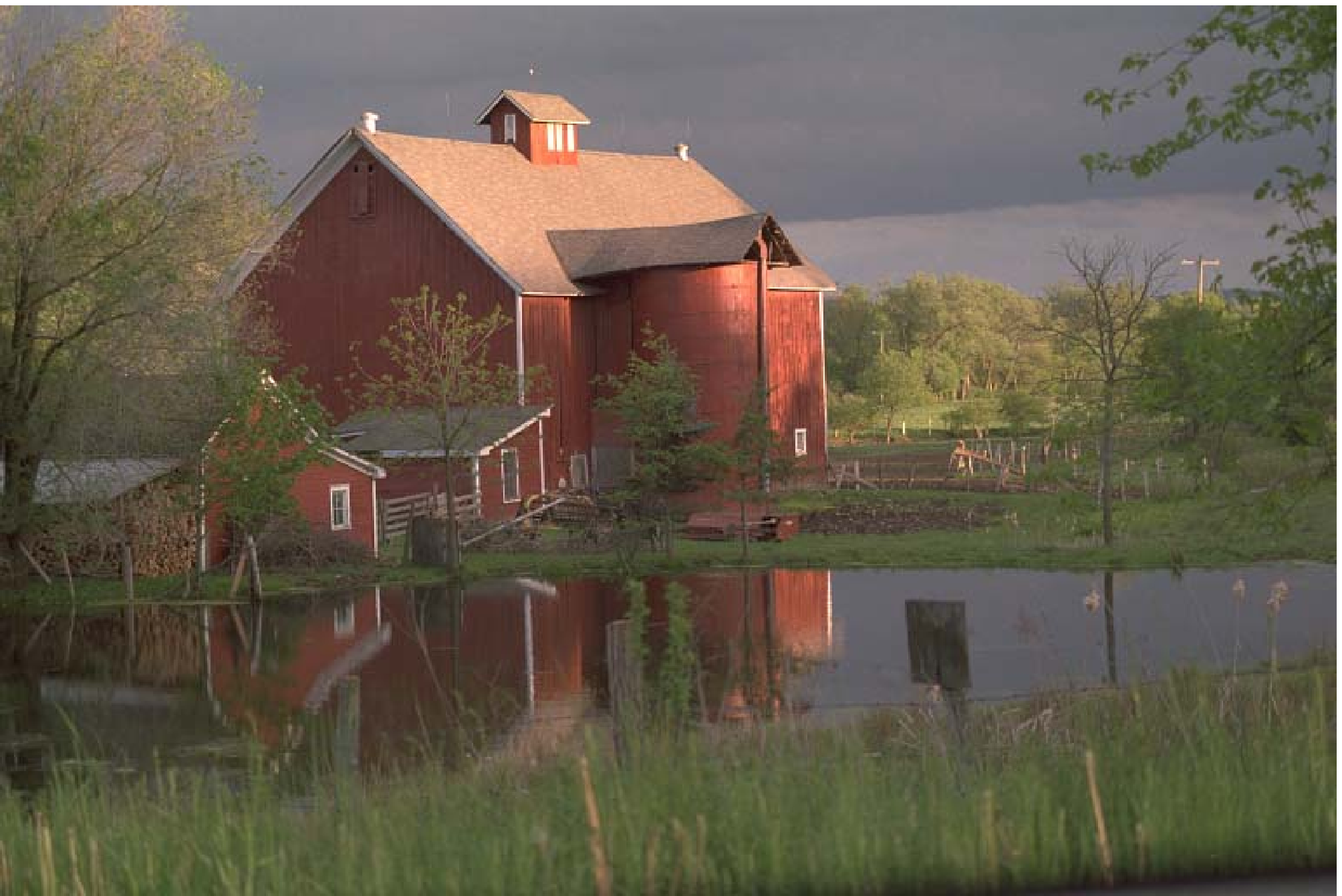}%
    \label{sfig:img3}%
   }%
   \hfil%
  \subfigure[Image 4]{%
   \includegraphics[width=.475\columnwidth]{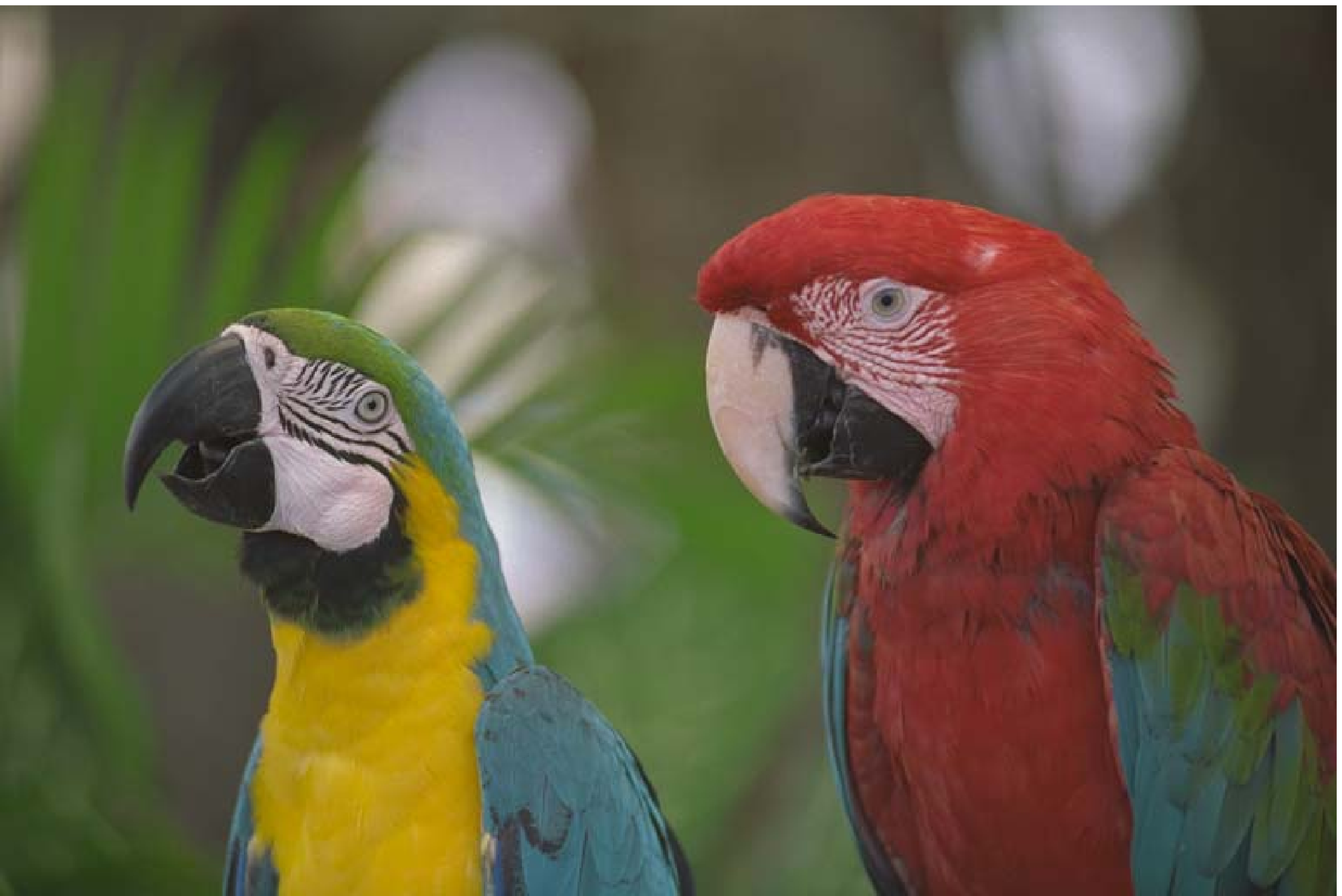}%
   \label{sfig:img4}%
  }%
 \caption{Test images.}
 \label{fig:test_image}
 \end{figure*}
 
  \begin{figure*}[t]
 \centering

  \subfigure[Image 1]{%
    \includegraphics[width=.475\columnwidth]{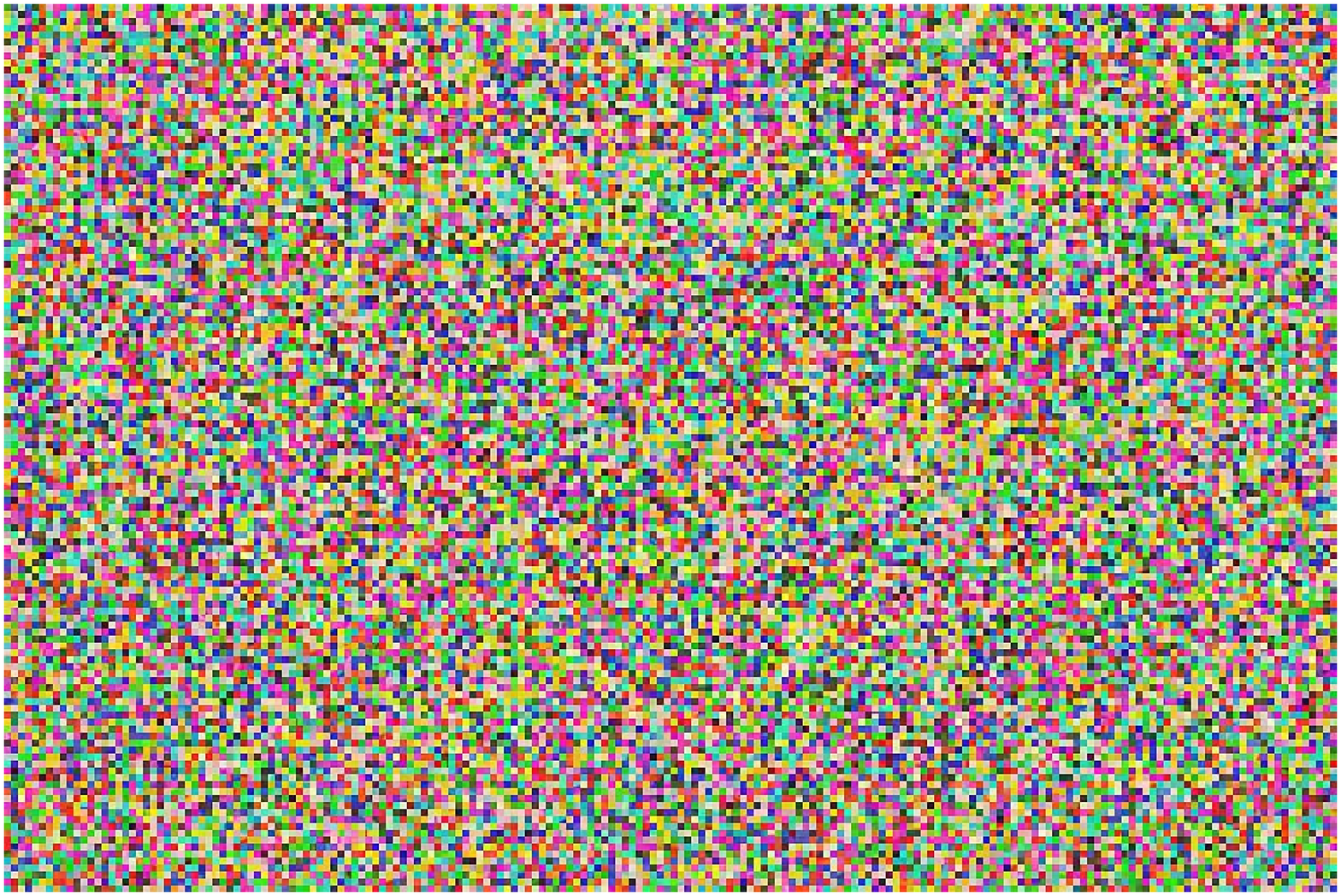}%
    \label{sfig:o_img1}%
   }%
   \hfil%
  \subfigure[Image 2]{%
   \includegraphics[width=.475\columnwidth]{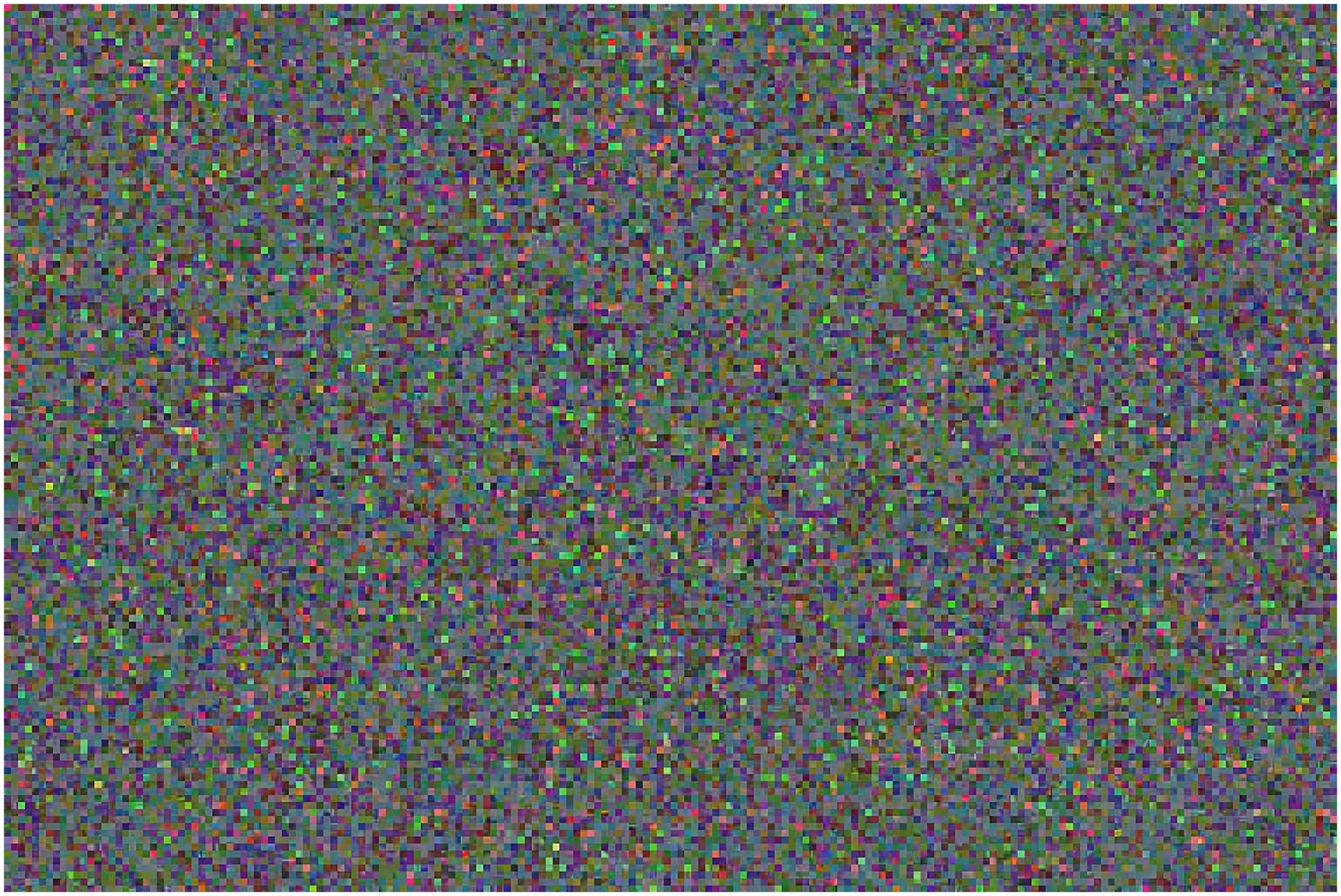}%
   \label{sfig:o_img2}%
   }%
   \hfil%
  \subfigure[Image 3]{%
    \includegraphics[width=.475\columnwidth]{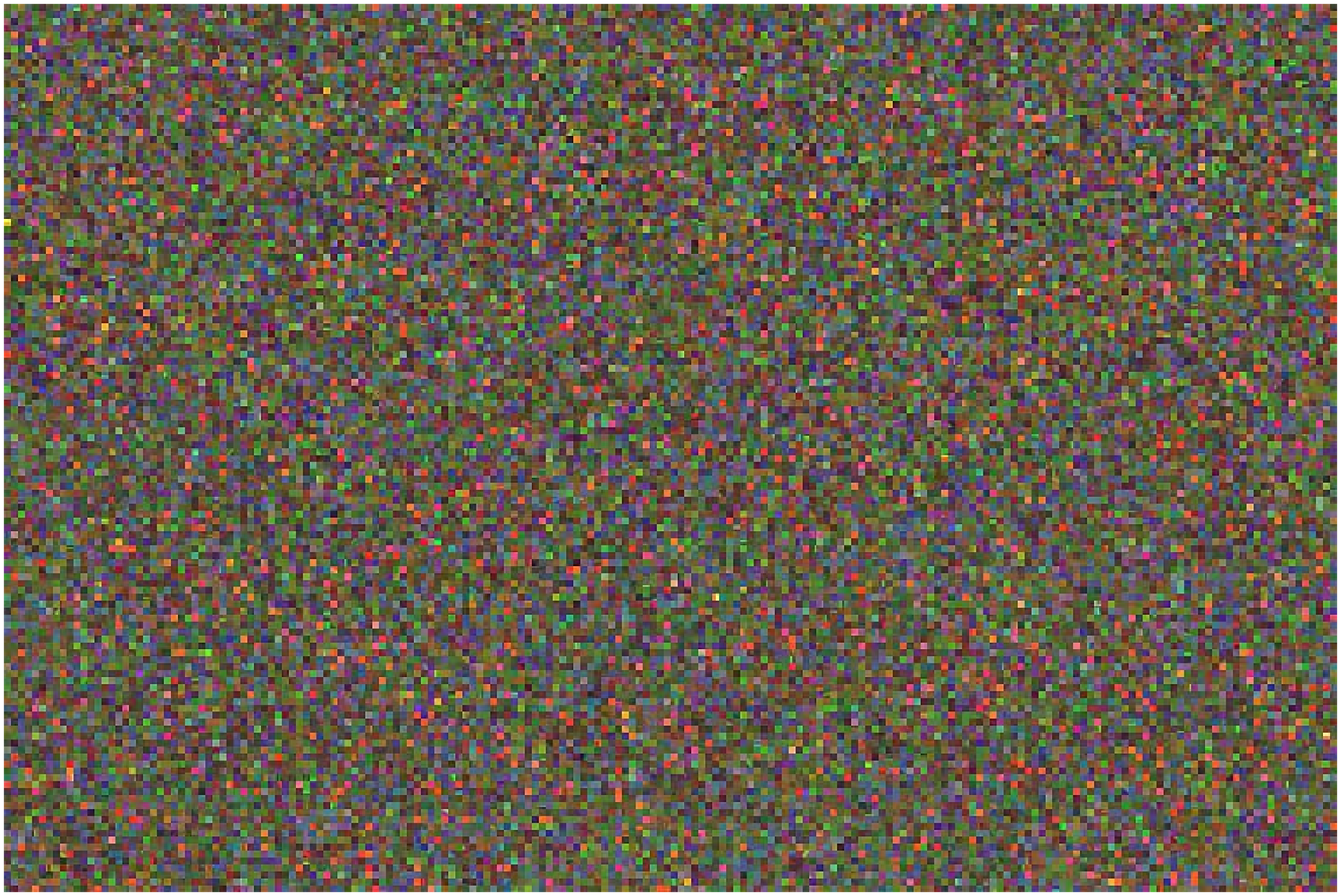}%
    \label{sfig:o_img3}%
   }%
   \hfil%
  \subfigure[Image 4]{%
   \includegraphics[width=.475\columnwidth]{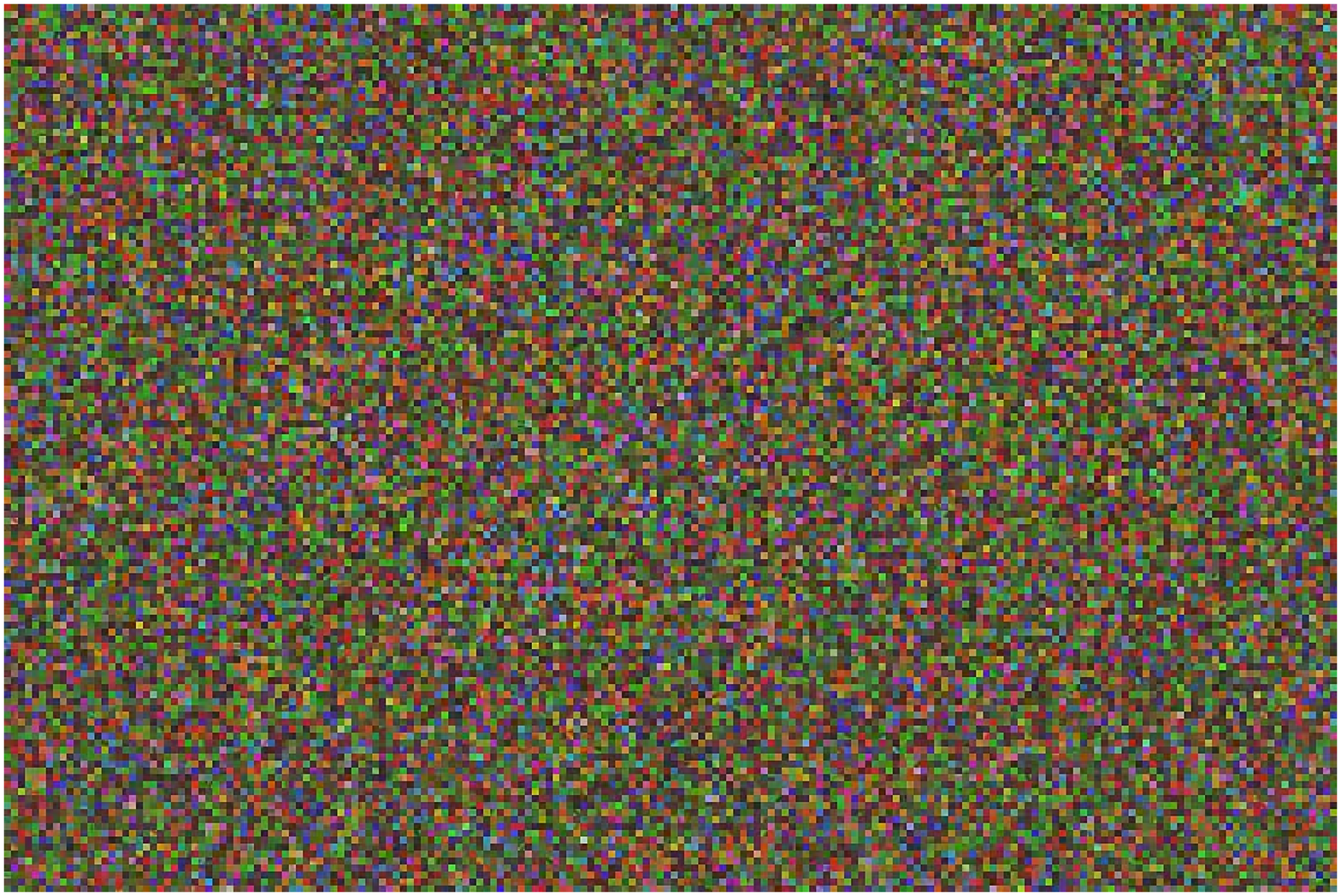}%
   \label{sfig:o_img4}%
  }%
 \caption{Output images by single-domain data hiding method (block size: $16 \times 16$ pixels).}
 \label{fig:output_image}
 \end{figure*}
 
   \begin{figure*}[t]
 \centering

  \subfigure[Image 1]{%
    \includegraphics[width=.475\columnwidth]{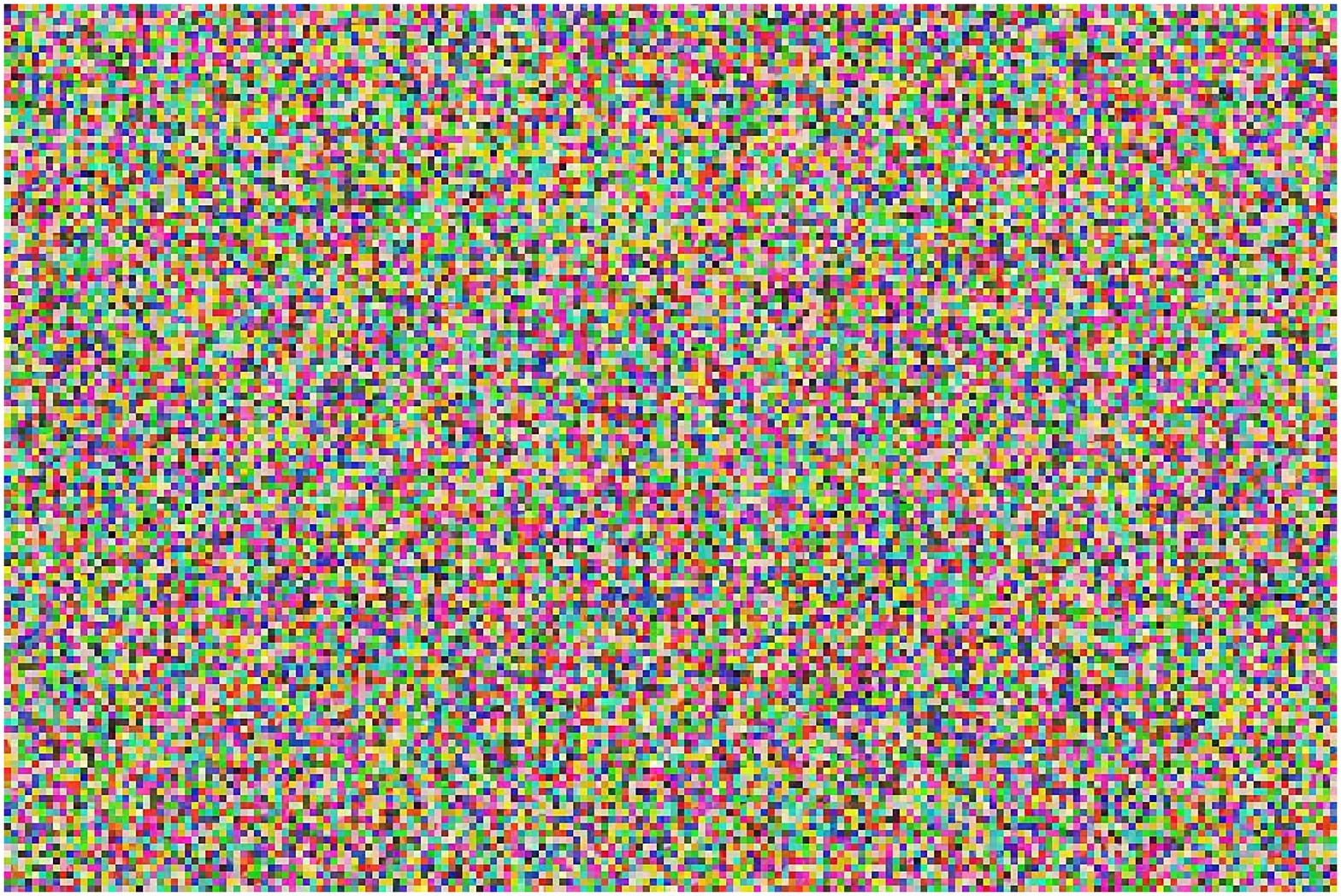}%
    \label{sfig:o_img1_ex}%
   }%
   \hfil%
  \subfigure[Image 2]{%
   \includegraphics[width=.475\columnwidth]{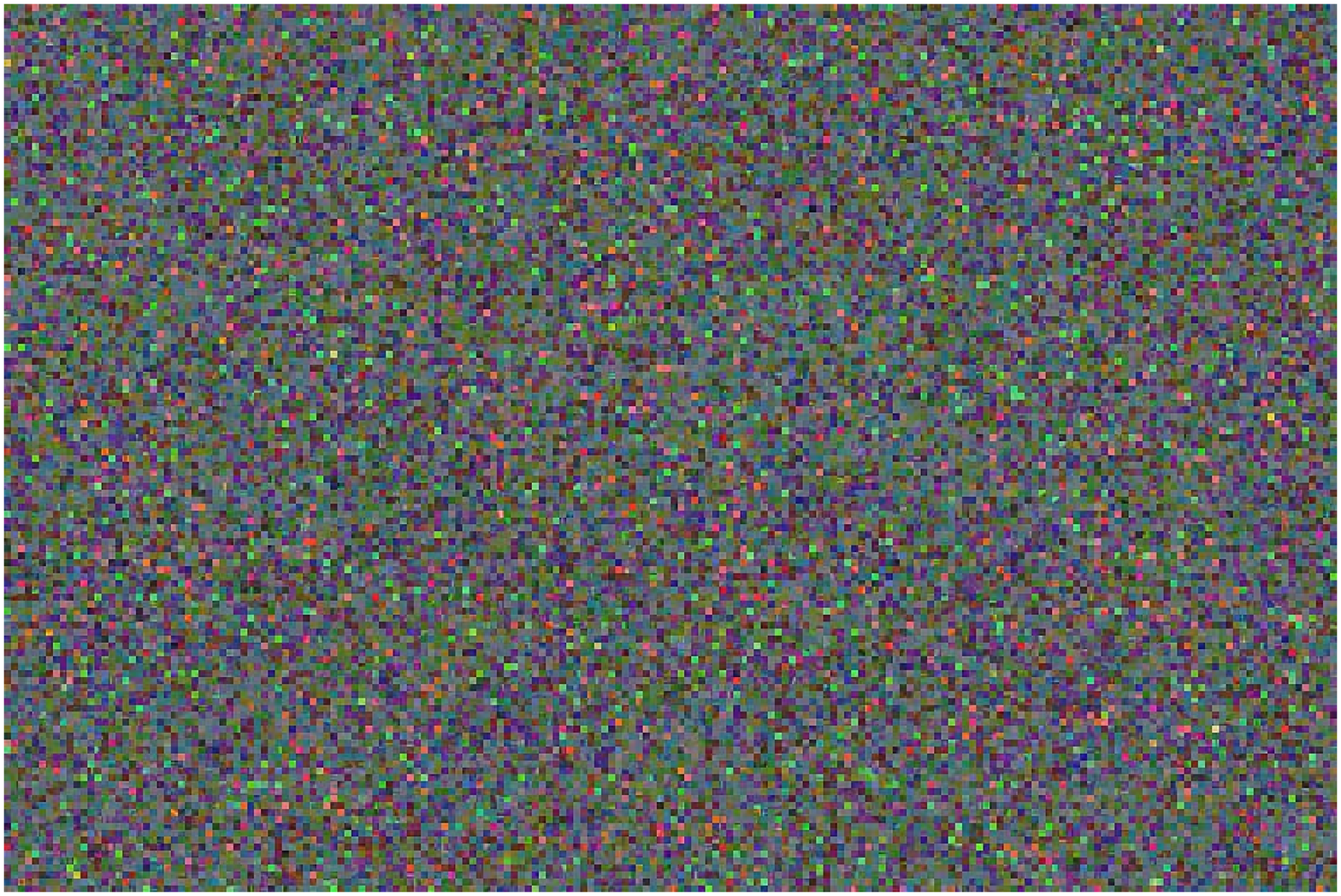}%
   \label{sfig:o_img2_ex}%
   }%
   \hfil%
  \subfigure[Image 3]{%
    \includegraphics[width=.475\columnwidth]{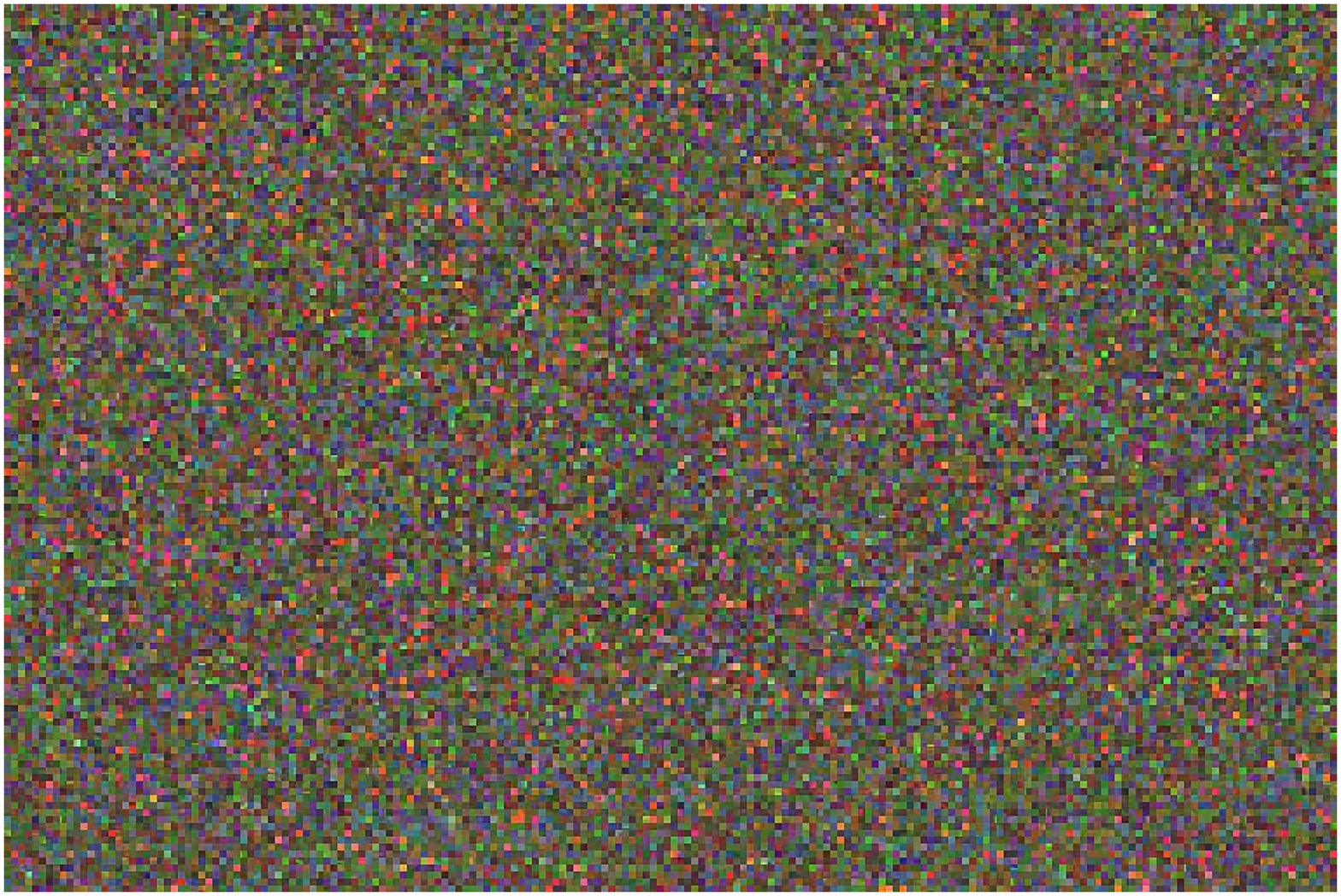}%
    \label{sfig:o_img3_ex}%
   }%
   \hfil%
  \subfigure[Image 4]{%
   \includegraphics[width=.475\columnwidth]{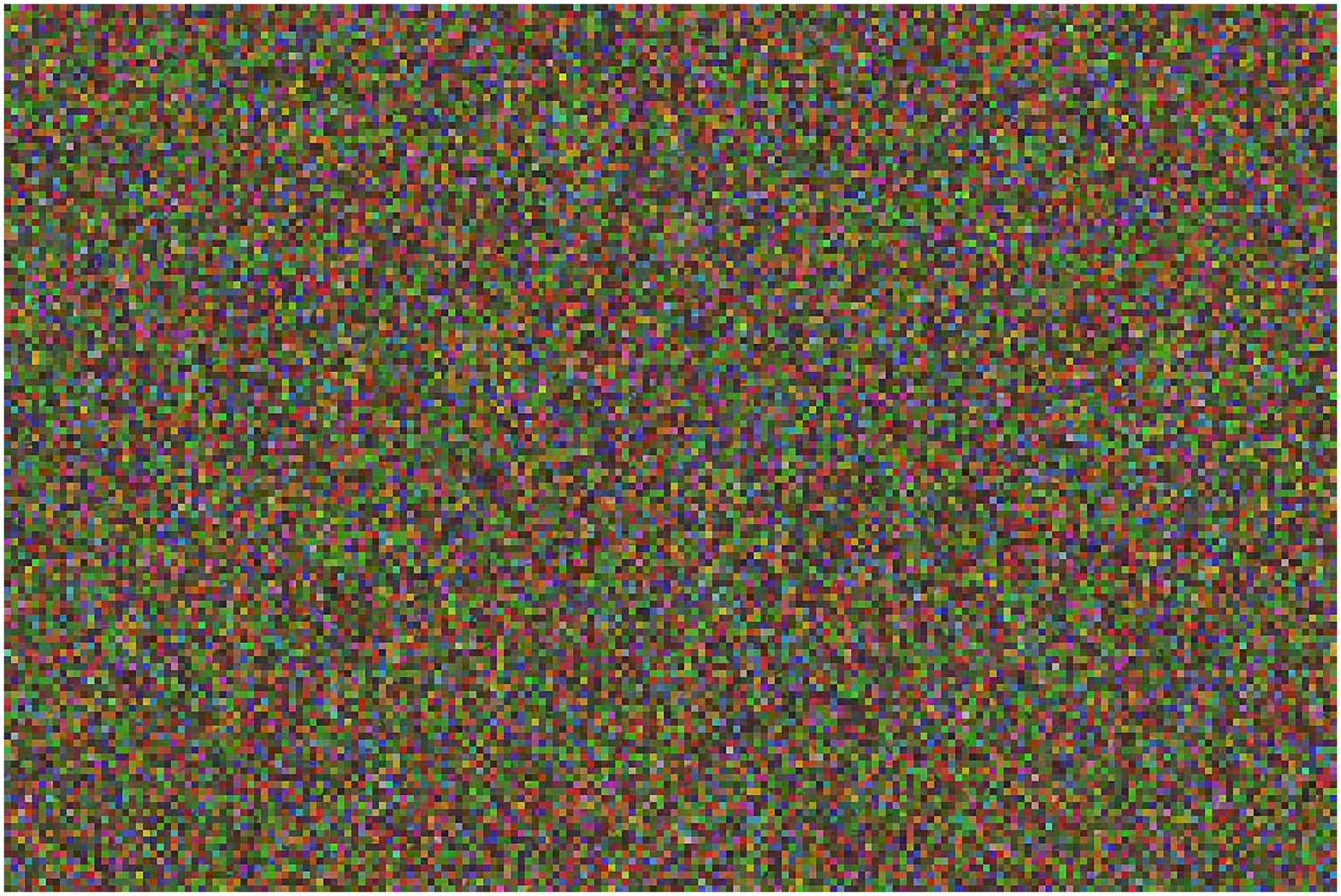}%
   \label{sfig:o_img4_ex}%
  }%
 \caption{Output images by two-domain data hiding method (block size: $16 \times 16$ pixels).}
 \label{fig:output_image_ex}
 \end{figure*}
 
 Here, we extend the proposed method to embed two different payloads in the plain and encrypted domains independently. 
 Figure~\ref{fig:flow_ex} shows the block diagram of the extended framework. 
 In this framework, we divide the original image into two regions beforehand. 
 Then, while one of the payloads is embedded into one region before encryption, the other payload can be embedded into the other region.
 We elaborate the procedure of this extension in the following steps. 
 \begin{description}
  \item{\bf{Step 1}:} Explore the $PP$ and $ZP$ in the histogram of the original image $I$. 
  \item{\bf{Step 2}:} Obtain the intermediate image $I'$ by shifting the histogram between the $PP$ and $ZP$. 
                      The shifted pixel values $X'$ are given by Eq.~\eqref{eq:hs1}. 
  \item{\bf{Step 3}:} Divide the image $I'$ into multiple blocks with $b_{x} \times b_{y}$ pixels. 
  \item{\bf{Step 4}:} Prepare two regions $A$ and $B$, and assign each block to either $A$ or $B$ using a pseudo-random sequence.
                      The $\alpha$-th block containing $x_{PP}$ assigned into $A$ and the $\beta$-th block containing $x_{PP}$ assigned into $B$ are represented as $B_{x_{PP}, A}(\alpha)$ and $B_{x_{PP}, B}(\beta)$, respectively.
  \end{description}
 For the blocks $B_{x_{PP}, A}(\alpha)$, 
 \begin{description}
  \item{\bf{Step 5A}:} Determine the data hiding order and the target blocks for encryption according to Steps 4 -- 7 in \ref{ssec:3-1}. 
  \item{\bf{Step 6A}:} Embed a payload into the pixels $x_{PP}$ in sequence. 
                       The marked pixel values $PP'$ are given by Eqs.~\eqref{eq:hs2} and \eqref{eq:hs3}.
  \item{\bf{Step 7A}:} Perform encryption for the target blocks. 
  \item{\bf{Step 8A}:} Obtain the output region $A'_{E}$.
 \end{description}
 Similarly, for the blocks $B_{x_{PP}, B}(\beta)$,  
 \begin{description}
  \item{\bf{Step 5B}:} Determine the data hiding order and the target blocks for encryption according to Steps 4 -- 7 in \ref{ssec:3-1}. 
  \item{\bf{Step 6B}:} Perform encryption for the target blocks. 
  \item{\bf{Step 7B}:} Embed another payload into the pixels $x_{PP}$ in sequence. 
                       The marked pixel values $PP'$ are given by Eqs.~\eqref{eq:hs2} and \eqref{eq:hs3}.
  \item{\bf{Step 8B}:} Obtain the output region $B'_{E}$.
 \end{description}
 Finally,
 \begin{description}
  \item{\bf{Step 9}:} By concatenating the regions $A'_{E}$ and $B'_{E}$, obtain the output image $I'_{E}$.
 \end{description}

 In this framework, the two independent regions are derived before the main processes. 
 Accordingly, the types of user authorities can be extended, e.g., data extraction in region $A$ only and data extraction in region $B$ with decryption. 

\section{Experimental results}
\label{sec:4}

 We specify the effectiveness of the proposed method from the aspects of lossless compression performance using JPEG-LS~\cite{JPEGLS} and JPEG 2000~\cite{JP2}, hiding capacity/image quality, and robustness against COAs. 
 The four $2,048 \times 3,072$ images~\cite{Kodak} shown in Fig.~\ref{fig:test_image} were used as test images. 
 The block size in our experiments is $16 \times 16$ pixels. 
 We tested 20 times for each image, namely, we generated 20 output images each. 
 Pseudo-random number sequences are used as the payload, and the payload amount is equal to the data hiding capacity. 
 Figure \ref{fig:output_image} shows the output images obtained by the proposed method with data hiding in a single domain. 
 Similarly, Fig.~\ref{fig:output_image_ex} depicts the output images by the extended proposed method for data hiding in two domains, as described in \ref{ssec:3-3}. 
 The output images obtained by those two methods are quite similar to each other. 
 In this experiment, we adopt the independent CE processing of RGB components~\cite{IEICE-T2018:SImaizumi} in the encryption process.
  
\subsection{Compression performance}
\label{ssec:4-3}
 \begin{figure*}[t]
 \centering

  \subfigure[Image 1]{%
    \includegraphics[width=.5\columnwidth]{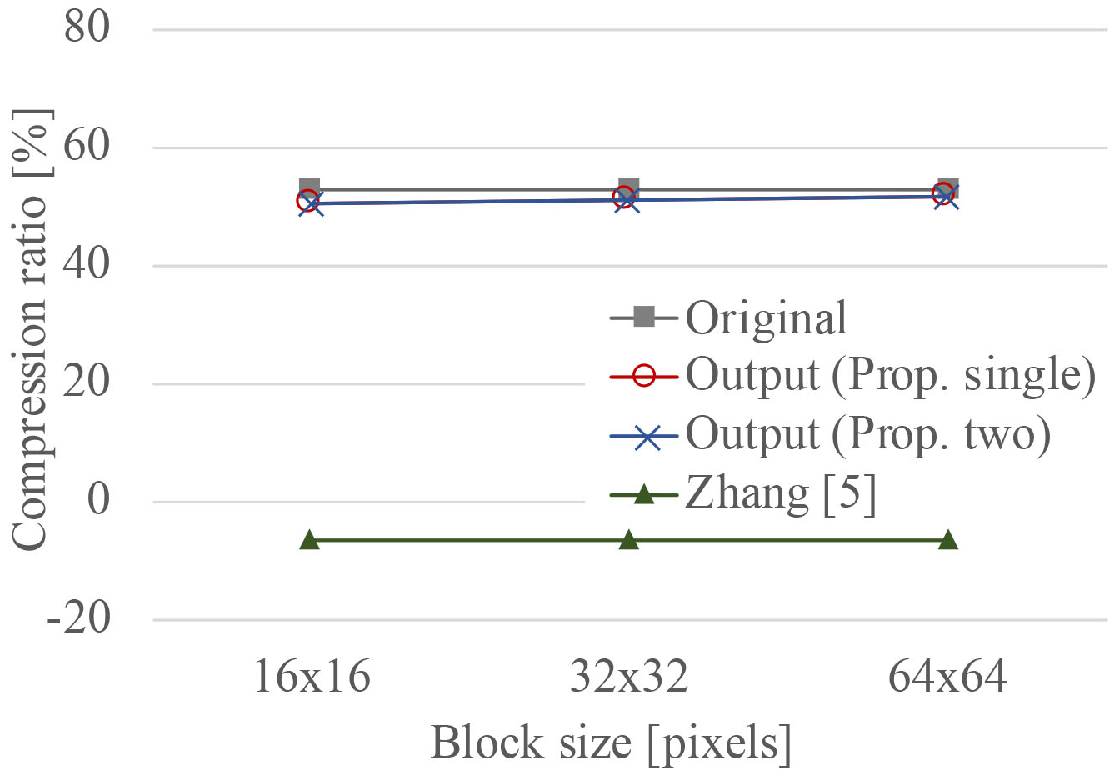}%
%    \label{sfig:prophist_air}%
   }%
   \hfil%
  \subfigure[Image 2]{%
   \includegraphics[width=.5\columnwidth]{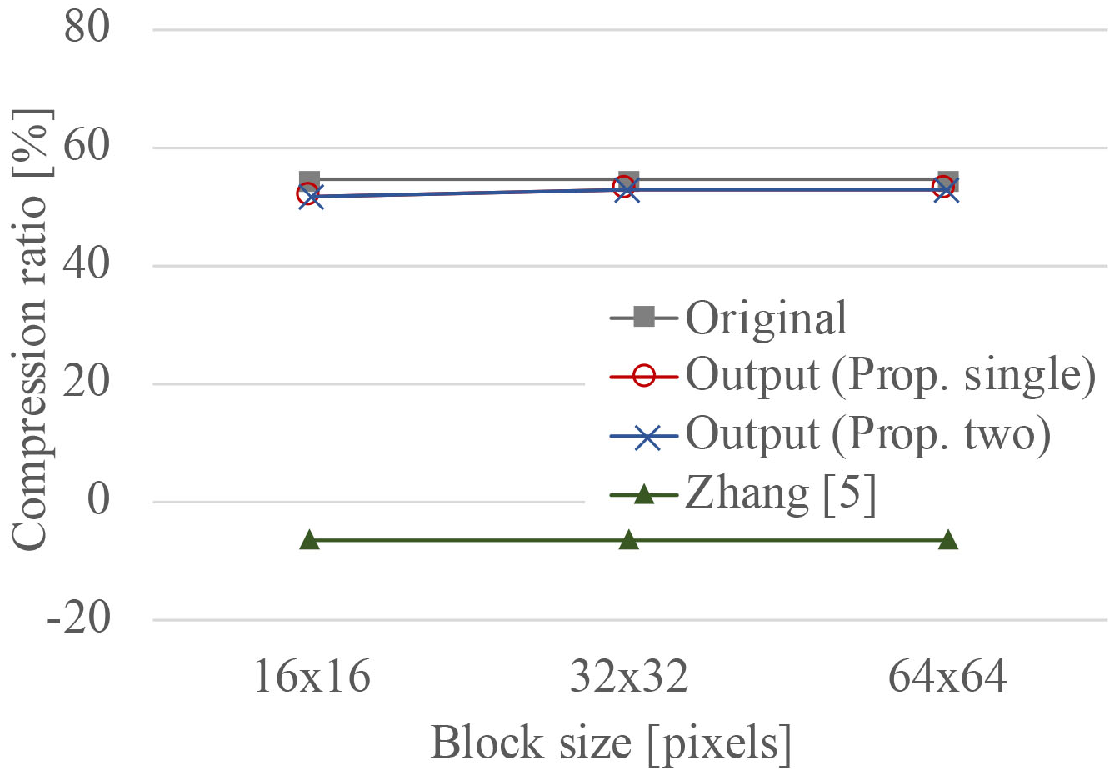}%
%   \label{sfig:convhist_air}%
   }%
   \hfil
  \subfigure[Image 3]{%
   \includegraphics[width=.5\columnwidth]{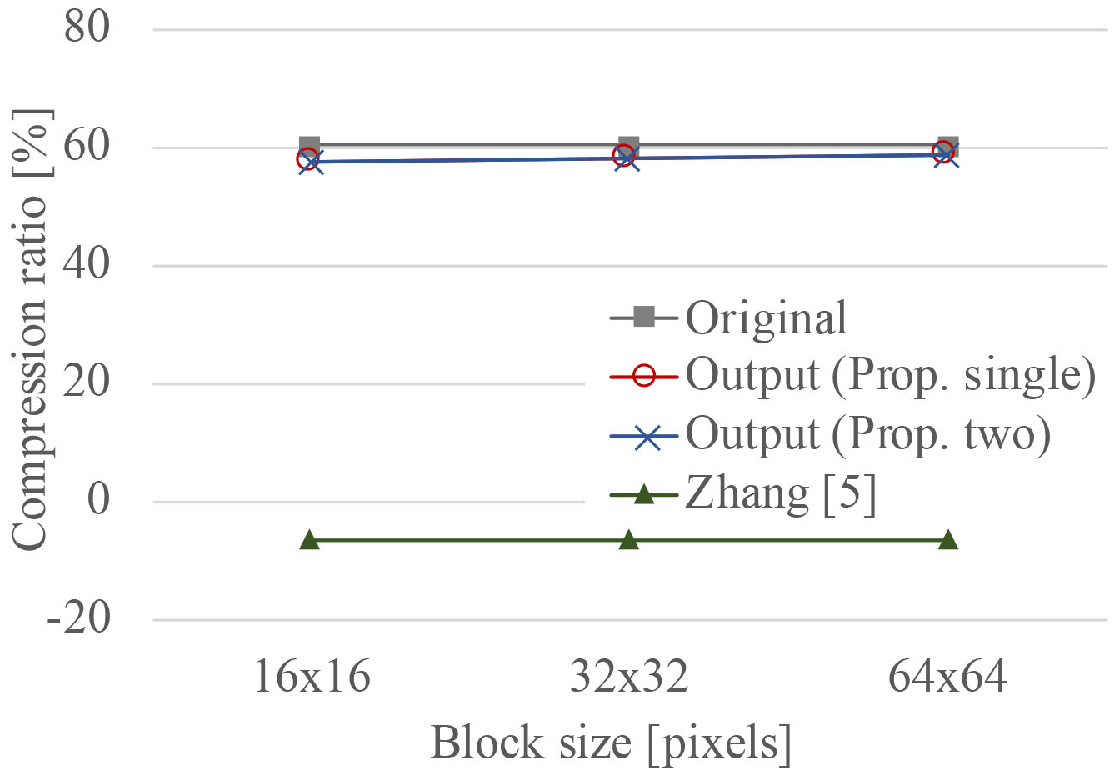}%
%   \label{sfig:orighist_air}%
   }%
   \hfil%
  \subfigure[Image 4]{%
   \includegraphics[width=.5\columnwidth]{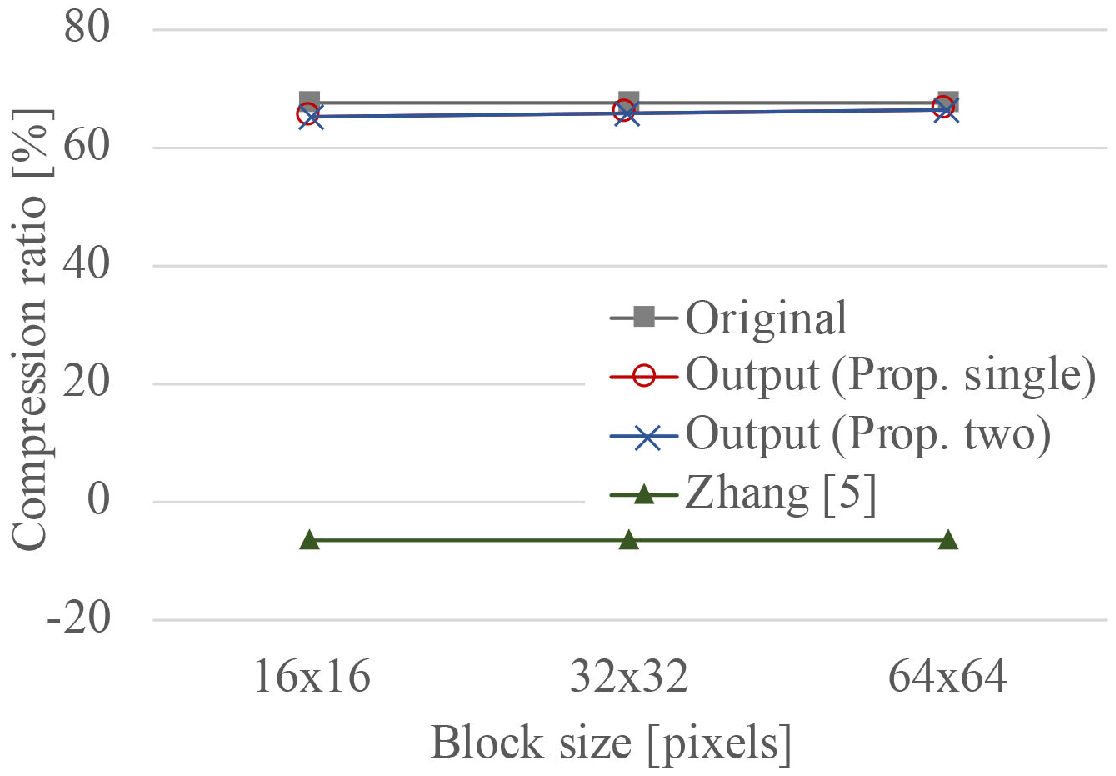}%
%   \label{sfig:convhist_air}%
  }% 
 \caption{Lossless compression performance using JPEG-LS.}
 \label{fig:JLScompress}
 \end{figure*}

 \begin{figure*}[t]
 \centering

  \subfigure[Image 1]{%
    \includegraphics[width=.5\columnwidth]{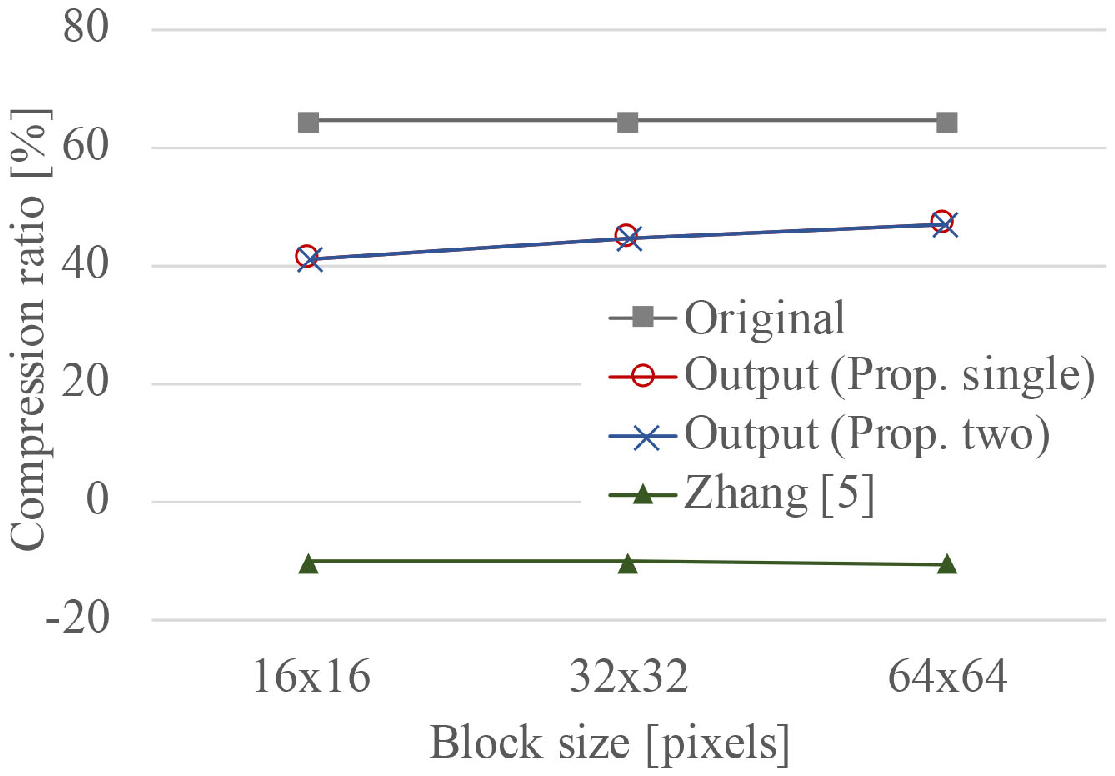}%
%    \label{sfig:prophist_air}%
   }%
   \hfil%
  \subfigure[Image 2]{%
   \includegraphics[width=.5\columnwidth]{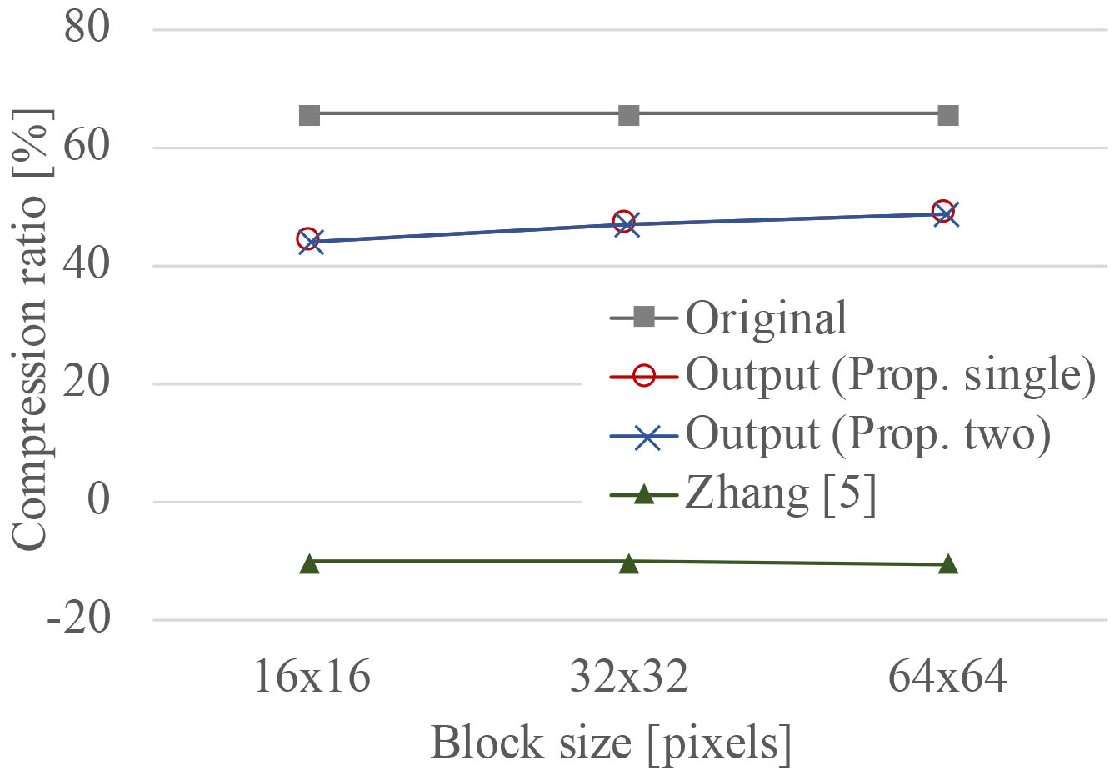}%
%   \label{sfig:convhist_air}%
   }%
   \hfil
  \subfigure[Image 3]{%
   \includegraphics[width=.5\columnwidth]{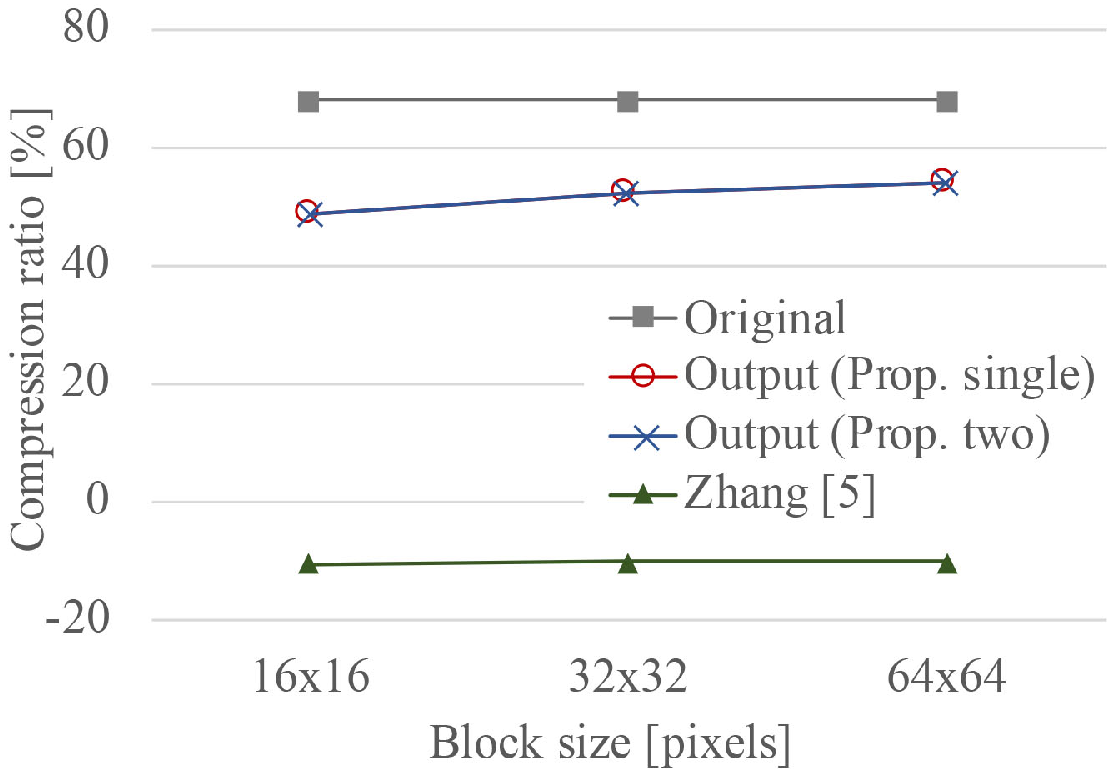}%
%   \label{sfig:orighist_air}%
   }%
   \hfil%
  \subfigure[Image 4]{%
   \includegraphics[width=.5\columnwidth]{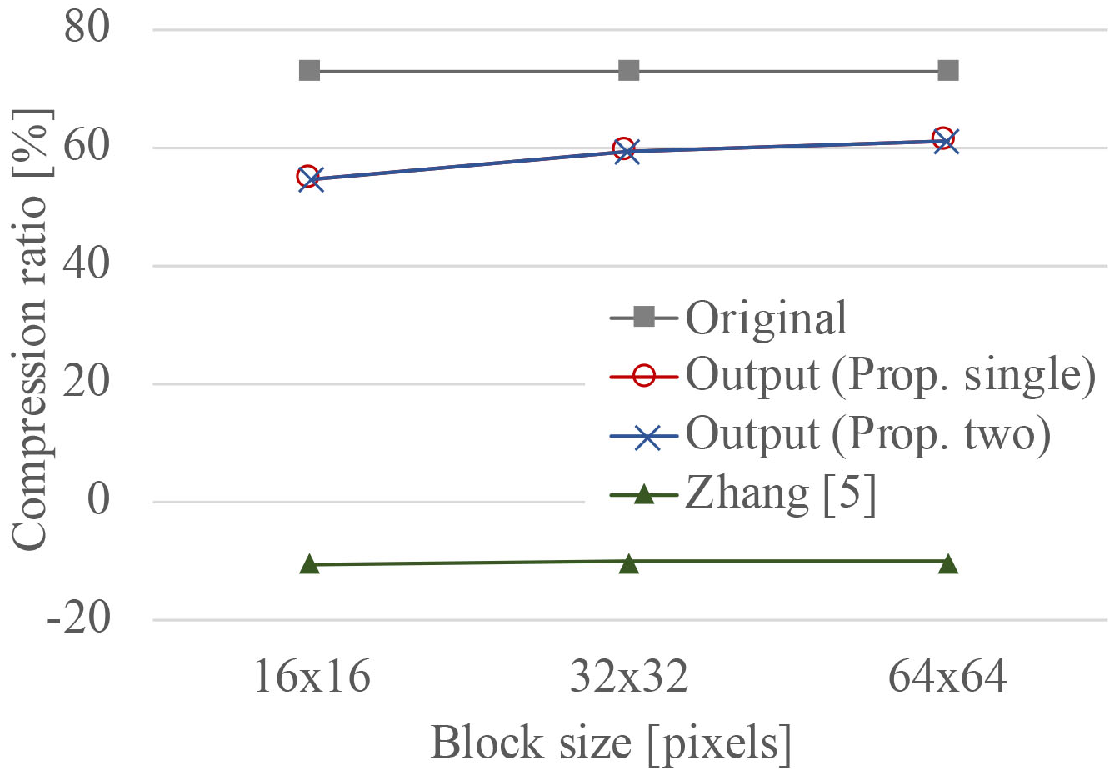}%
%   \label{sfig:convhist_air}%
  }% 
 \caption{Lossless compression performance using JPEG 2000.}
 \label{fig:JP2compress}
 \end{figure*}

 We evaluate the lossless compression performance using JPEG-LS and JPEG 2000. 
 Figures~\ref{fig:JLScompress} and \ref{fig:JP2compress} show the compression ratio of the original and output images, where the block size is $16 \times 16$, $32 \times 32$, or $64 \times 64$ pixels.
 Those values are the average of the compression ratio for 20 output images. 
 It is confirmed that the output images by the proposed methods can be greatly compressed while those obtained by the pixel-based encryption method~\cite{IEEE-SPL2011:XZhang} cannot be compressed at all. 
 According to Fig.~\ref{fig:JLScompress}, the original and output images show quite analogous results in JPEG-LS compression. 
 In contrast, the CE process marginally affects the compression ratio of the output images using JPEG 2000, as shown in Fig.~\ref{fig:JP2compress}. 
 This is because a discrete wavelet transform in the JPEG 2000 coding system uses the correlation calculated from the wider spatial range of an image than JPEG-LS. 
 In our methods, the compression performance is better when the block size is larger.

\subsection{Data hiding capacity and image quality}
\label{ssec:4-2}
 The data hiding capacity and the marked-image quality of the proposed methods are compared with those of Zhang's method~\cite{IEEE-SPL2011:XZhang}. 
 The marked image means the decryption-only image here. 
 Table~\ref{tb:capacity_and_quality} shows the comparison results. 
 The proposed methods are superior to Zhang's method in both the capacity and the image quality.  
 In Zhang's method, one bit is embedded into each divided block, and thus the hiding capacity depends on the block size. 
 In this experiment, the block size in Zhang's method is set as $16 \times 16$ pixels.
 If the block size is smaller than $16 \times 16$ pixels, the data hiding capacity increases, but the extracted-bit error rate becomes higher. 
 Even when the block size is $16 \times 16$ pixels, the payload cannot be extracted correctly in those test images. 
 Additionally, the lower three bits are flipped in half of all the pixels statistically, and thus the total number of flipped bits directly affects the marked-image quality. 
 In contrast, in the proposed methods, both the data hiding capacity and the marked-image quality are constant irrespective of block size.
 
 \begin{table*}[t]
  \begin{center}
  \caption{Data hiding capacity and marked-image quality}
  \label{tb:capacity_and_quality}
    \begin{tabular}{|l|c|c|c|c|c|c|c|c|} \hline 
                                      & \multicolumn{2}{c|}{Image 1} & \multicolumn{2}{c|}{Image 2} & \multicolumn{2}{c|}{Image 3}& \multicolumn{2}{c|}{Image 4} \\ \hline 
     Method                           & Capacity [bits] & PSNR [dB] & Capacity [bits] & PSNR [dB] & Capacity [bits] & PSNR [dB] & Capacity [bits] & PSNR [dB] \\ \hhline{|=|=|=|=|=|=|=|=|=|}
     Prop. single                     & 313,482         & 57.44     & 601,220         & 52.56     & 484,534         & 52.00     & 528,158         & 54.55     \\ \hline
     Prop. two                        & 313,482         & 57.44     & 601,220         & 52.56     &484,534          & 52.00     &528,158          & 54.55     \\ \hline
     Zhang~\cite{IEEE-SPL2011:XZhang} & 73,728          & 41.52     & 73,728          &41.50      &73,728           &41.49      &73,728           &41.50      \\ \hline
   \end{tabular}
  \end{center}
 \end{table*}

\subsection{Robustness against ciphertext-only attacks}
\label{ssec:4-1}
 \begin{figure}[t]
 \centering

  \subfigure[Original]{%
    \includegraphics[width=.24\columnwidth]{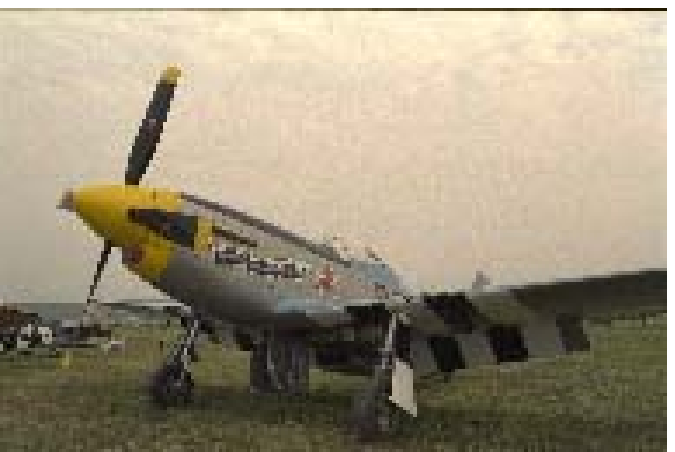}%
    \label{sfig:mini_original}%
   }%
   \hfil%
  \subfigure[Output (Prop. single)]{%
   \includegraphics[width=.24\columnwidth]{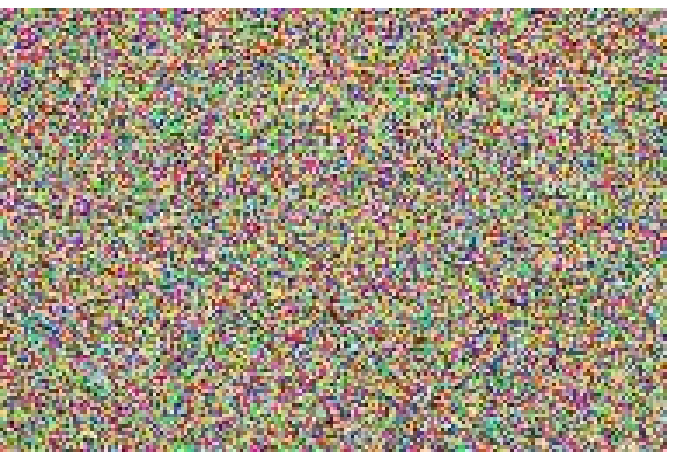}%
   \label{sfig:mini_prop}%
   }%
 \caption{Resized images.}
 \label{fig:resized_image}
 \end{figure}

 \begin{table}[t]
  \begin{center}
  \caption{Correlation coefficients between neighboring pixels in resized image.}
  \label{tb:correlation_coefficient}
   \begin{tabular}{|c|l|c|c|c|} \hline
     \multicolumn{2}{|c|}{}  & Horizontal & Vertical & Diagonal \\ \hhline{|=|=|=|=|=|}
            & Original              &  0.9776 &  0.9539 &  0.9432  \\ \cline{2-5}
    Image 1 & Output (Prop. single) & -0.0024 &  0.0012 &  0.0010  \\ \cline{2-5}
            & Output (Prop. two)    & -0.0077 &  0.0009 & -0.0003  \\ \hline
            & Original &  0.8186    &  0.7534 &  0.7110  \\ \cline{2-5}
    Image 2 & Output (Prop. single) & -0.0017 & -0.0024 & -0.0013  \\ \cline{2-5}
            & Output (Prop. two)    &  0.0002 &  0.0014 &  0.0051  \\ \hline
            & Original &  0.8221    &  0.8953 &  0.8348  \\ \cline{2-5}
    Image 3 & Output (Prop. single) & -0.0016 & -0.0025 & -0.0010  \\ \cline{2-5}
            & Output (Prop. two)    & -0.0022 & -0.0058 &  0.0077  \\ \hline
            & Original &  0.9378    &  0.9366 &  0.9298   \\ \cline{2-5}
    Image 4 & Output (Prop. single) &  0.0016 & -0.0010 &  0.0015   \\ \cline{2-5}
            & Output (Prop. two)    & -0.0260 &  0.0049 &  0.0016   \\ \hline
   \end{tabular}
  \end{center}
 \end{table}

 Here, we consider robustness against COAs, where an attacker is assumed to have access only to ciphertexts. 
 Our CE method is based on the premise that encryption keys are securely maintained, and the CE method prepares different encryption keys for each image/user. 
 Brute force attacks and jigsaw puzzle solver (JPS) attacks are cited as COAs for the CE method. 
 The robustness against those two types of attacks has already been evaluated in our previous works such as~\cite{IEICE-T2015:KKurihara, IEICE-T2018:SImaizumi, IEICE-T2017:KKurihara, IEICE-T2018:TChuman, APSIPA-T2019:WSirichotedumrong}. 
 The robustness does not deteriorate even when a CE image contains a payload.
 JPS attacks prompt the assembly of a jigsaw puzzle by using the correlation among pieces. 
 We regard blocks of a CE image as pieces of a jigsaw puzzle. 
 Although the robustness against JPS attacks has been revealed by our previous works, we purposely compute the correlation coefficient~\cite{McGraw-Hill2018:AGBluman} to confirm a part of the correlation among blocks. 
 The correlation coefficient has been used in multiple literatures~\cite{OptComm2009:CKHuang, SignalProcess2016:ABelazi} for security analysis. 

 The output image is encrypted block by block, and thus the correlation among the neighboring blocks is calculated. 
 In EtC images, the correlation among pixels within each block is retained for high-performance compression. 
 As shown in Fig.~\ref{fig:resized_image}, we derived the resized images of the original and output images by taking the top-left pixels from each block, where the image size is reduced to $128 \times 198$ pixels.
 
 The 2,000 pairs of the neighboring pixels are randomly chosen in horizontal, vertical, and diagonal directions. 
 Then, the correlation coefficients are calculated for each direction. 
 The correlation coefficient $r_{x,y}$ between the two neighboring pixels $x$ and $y$ is given as 
 \begin{equation}
  r_{x,y} = \frac{cov(x,y)}{\sqrt{D(x)}\sqrt{D(y)}}, 
  \label{eq:r_xy}
 \end{equation}
 where $D(x)$, $D(y)$, and $cov(x, y)$ are represented as 
 \begin{eqnarray}
	D(x) = &  {\frac{1}{S}}\sum_{j=1}^{S}(x_j-E(x))^2, \label{eq:Dx} \\
	D(y) = &  {\frac{1}{S}}\sum_{j=1}^{S}(y_j-E(y))^2, \label{eq:Dy} \\
	cov(x,y) = &  {\frac{1}{S}}\sum_{j=1}^{S}(x_j-E(x))(y_j-E(y)). \label{eq:cov-xy}
 \end{eqnarray}
 Here, $S$ represents the number of neighboring-pixel pairs ($S=2,000$ in this experiment), and $E(x)$ and $E(y)$ are the average of the pixels $x_i$ and $y_i$, which are shown as
 \begin{eqnarray}\hspace{12pt}\vspace{-7pt}
	E(x) = &  {\frac{1}{S}}\sum_{j=1}^{S}x_j \label{eq:Ex}, \\
    E(y) = &  {\frac{1}{S}}\sum_{j=1}^{S}y_j \label{eq:Ey}.
 \end{eqnarray}
 Table~\ref{tb:correlation_coefficient} shows the correlation coefficients $r_{x,y}$ for the original and output images. 
 Those values are the average of $r_{x,y}$ for 20 output images.
 It is verified that the $r_{x,y}$ values of the output images are close to 0, and thus the correlation among blocks is quite low.

\section{Conclusions}
\label{sec:5}
 We proposed an effective RDH method that embeds a payload in the encrypted domain and can directly extract the payload from the decrypted image. 
 In an opposite fashion, the proposed method can also embed a payload in the plain domain and extract the payload from the encrypted image. 
 Namely, we can choose either the plain or encrypted domain for data hiding. 
 The compressible encryption method for the EtC system is adopted in the proposed method, and the HS-based RDH method is integrated into our framework. 
 We further extended the proposed method to embed a payload in the plain and encrypted domains.
 The output images obtained by the proposed methods have been evaluated in terms of lossless compression performance by JPEG-LS and JPEG 2000, data hiding capacity/marked-image quality, and correlation among blocks. 
 Our new method can provide an efficient framework to flexibly integrate both encryption and data hiding techniques. 
 
\section*{Acknowledgement}
This work was partially supported by Grant-in-Aid for Research Activity start-up, No.19K23070, from the Japan Society for the Promotion Science.

%\end{document}

\profile{Shoko IMAIZUMI}{received her B. Eng., M. Eng., and Ph.D. degrees from Tokyo Metropolitan University, Japan in 2002, 2005, and 2011. 
In 2011, she joined Chiba University, where she is currently an Associate Professor of the Graduate School of Engineering. 
From 2003 to 2004, she was with the Ministry of Education, Culture, Sports, Science and Technology of Japan. 
She was a Researcher at the Industrial Research Institute of Niigata Prefecture from 2005 to 2011. 
Her research interests include image processing and multimedia security. 
She served as an Associate Editor for IEICE Trans. Fundamentals in 2016-2020, and is currently a Director for SPIJ (Society of Photography and Imaging of Japan). 
She is a member of IEICE, ITE, IEEJ, SPIJ, IEEE, and APSIPA.}
\profile{Yusuke IZAWA}{received his B. Eng. and M. Eng. degrees from Chiba University, Japan in 2018 and 2020. 
He joined the National Printing Bureau in 2020. 
His research interests include multimedia security.}
\profile{Ryoichi HIRASAWA}{received his B.Eng. degree from Chiba University, Japan in 2019. Since 2019, he has been a Master course student at Chiba University. His research interests include image processing.}
\profile{Hitoshi KIYA}{received his B.E. and M.E. degrees from Nagaoka University of Technology, Japan in 1980 and 1982 and his Dr. Eng. degree from Tokyo Metropolitan University in 1987. In 1982, he joined Tokyo Metropolitan University, where he became Full Professor in 2000. From 1995 to 1996, he attended the University of Sydney, Australia, as a Visiting Fellow. He is a Fellow of IEEE, IEICE, and ITE. He currently serves as President of APSIPA, and he served as Inaugural Vice President (Technical Activities) of APSIPA in 2009-2013 and as Regional Director-at-Large for Region 10 of the IEEE Signal Processing Society in 2016-2017. He was also President of the IEICE Engineering Sciences Society in 2011-2012, and he served there as Vice President and Editor-in-Chief for the IEICE Society Magazine and Society Publications. He has been an Editorial Board Member of eight journals, including IEEE Trans. on Signal Processing, Image Processing, and Information Forensics and Security, Chair of two technical committees, and Member of nine technical committees including the APSIPA Image, Video, and Multimedia Technical Committee (TC) and IEEE Information Forensics and Security TC. He has organized a lot of international conferences in such roles as TPC Chair of IEEE ICASSP 2012 and as General Co-Chair of IEEE ISCAS 2019. Dr. Kiya has received numerous awards, including ten best paper awards.}

%%\profile*{}{}% without picture of author's face
%
\end{document}